%% file: main.tex
\documentclass[runningheads]{llncs}
\usepackage[T1]{fontenc}
\usepackage{graphicx}
\usepackage{tabularx} %
\usepackage{subcaption}
\usepackage{soul}
\usepackage{multirow}
\usepackage{wrapfig}
\usepackage[utf8]{inputenc}
\usepackage{microtype}  %
\usepackage{algorithm}
\usepackage{algorithmic}

\usepackage[mode=math,group-separator={,},group-minimum-digits=4]{siunitx}
\usepackage{nth}
\usepackage{longtable}
\usepackage{booktabs}
\usepackage{circledsteps}
\usepackage[nolist]{acronym}
\usepackage{xspace}
\usepackage{comment}
\usepackage{bm} %
\usepackage{todonotes}
\usepackage{arydshln} %
\usepackage{pifont} %
\usepackage{ifthen}
\usepackage{csquotes}
\usepackage{hyperref}
\usepackage[inline]{enumitem}
\usepackage{mathtools}
\usepackage{amsmath}
\usepackage{amssymb}
\usepackage{multicol}
\usepackage{appendix}
\usepackage[most]{tcolorbox} %
\usepackage{xcolor}          %
\usepackage{tikz}
\usetikzlibrary{shapes.geometric, arrows.meta, positioning}
\usepackage{circledtext}
\usepackage{ragged2e}
\usepackage{textgreek}
\usepackage{cleveref} %
\usepackage{pdflscape}
\usepackage{makecell}

\input{macros}
\input{acronyms}

\begin{document}

\title{Does Delegation Deliver? Identifying Challenges in Proof of Stake with Delegation}
\author{}
\authorrunning{S. Sokoto et al.}
\institute{\begin{tabular}{@{}l@{}}Saidu Sokoto, City St George's, Univ. of London, UK, saidu.sokoto@city.ac.uk\\
Leonhard Balduf, TU Darmstadt, Germany, leonhard.balduf@tu-darmstadt.de\\
Onur Ascigil, Lancaster University, UK, o.ascigil@lancaster.ac.uk\\
Wen Yang, HKUST (Guangzhou), China, wyang330@connect.hkust-gz.edu.cn\\
Qiming Ye, HKUST (Guangzhou), China, qiming@connect.hkust-gz.edu.cn\\
Gareth Tyson, HKUST (Guangzhou), China, gtyson@ust.hk\\
Micha{\l} Kr{\'o}l, City St George's, Univ. of London, UK, michal.krol@city.ac.uk\end{tabular}}
\titlerunning{Does Delegation Deliver?}
\maketitle

\begin{abstract}
Recently, multiple blockchains have adopted Proof of Stake with Delegation (PoSD), which allows open validator membership with voting power proportional to delegated stake. PoSD now secures blockchains worth $>$\$120B.
We argue that PoSD can only deliver on its promise if it satisfies 3 properties: power decentralization, sustainable economics and quality-based delegation.
We empirically measure these properties using Solana, the largest PoSD deployment, over a 15-month period.
We find that power remains highly concentrated for both validators and delegators, with 0.003\% of delegators controlling a blocking share of stake.
Economic sustainability is fragile, with up to 45\% of validators operating at a loss and high revenue concentration.
Finally, delegators respond far more strongly to commission and ecosystem affiliation than to validator quality or reliability.
We then provide recommendations for the ecosystem.

\end{abstract}

\input{sections/introduction}
\input{sections/background}

\input{sections/methodology}

\input{sections/centralization}

\input{sections/economics}

\input{sections/voting}

\input{sections/discussion}

\input{sections/related_work}

\newpage
\bibliographystyle{splncs04}
\bibliography{references}

\appendix
\input{appendix/appendix}

\end{document}

%% file: macros.tex
\newboolean{showcomments}
\setboolean{showcomments}{false}
\newcommand\important[1]{\todo[inline]{\textbf{Important:} #1}}
\newcommand\lb[1]{\textbf{\textcolor{orange}{Leo: #1}}}
\newcommand\michal[1]{\textbf{\textcolor{magenta}{Michal: #1}}}
\newcommand\onur[1]{\textbf{\textcolor{red}{Onur: #1}}}
\newcommand\saidu[1]{\textbf{\textcolor{blue}{Saidu: #1}}}
\newcommand\wen[1]{\textbf{\textcolor{purple}{Wen: #1}}}
\newcommand\qm[1]{\textbf{\textcolor{purple}{QM: #1}}}
\newcommand\gareth[1]{\textbf{\textcolor{brown}{GT: #1}}}

\ifthenelse{\boolean{showcomments}} { }
{
	\renewcommand\important[1]{\ignorespaces}
	\renewcommand\lb[1]{\ignorespaces}
	\renewcommand\michal[1]{\ignorespaces}
	\renewcommand\onur[1]{\ignorespaces}
	\renewcommand\saidu[1]{\ignorespaces}
	\renewcommand\wen[1]{\ignorespaces}
    \renewcommand\qm[1]{\ignorespaces}
    \renewcommand\gareth[1]{\ignorespaces}
}

\ifthenelse{\boolean{showcomments}}
{ \newcommand{\mynote}[3]{
		\protect\fbox{\bfseries\sffamily\scriptsize#1}
		{\small\textsf{\emph{\color{#3}{#2}}}}}}
{ \newcommand{\mynote}[3]{\ignorespaces}}

\newcommand{\etal}{\textit{et al.}\@\xspace}
\newcommand{\eg}{\textit{e.g.},\xspace}

\newcommand{\ie}{\textit{i.e.},\xspace}

\newcommand\para[1]{\vspace{0.05in} \noindent \textbf{#1.}}

\def\second{({\it ii})\xspace}

\definecolor{verylightgray}{gray}{0.8}

\crefname{section}{Sec.}{Sec.}
\Crefname{section}{Section}{Sections}
\crefname{equation}{eq.}{eq.}
\crefname{figure}{Fig.}{Fig.s}
\Crefname{figure}{Figure}{Figures}

\floatstyle{plain}
\newfloat{lstfloat}{htbp}{lop}
\floatname{lstfloat}{Listing}

\crefalias{lstfloat}{listing}

\usetikzlibrary{positioning}
\usetikzlibrary{graphs}
\usetikzlibrary{shapes.multipart}
\usetikzlibrary{shapes.misc}
\usetikzlibrary{matrix}
\usetikzlibrary{shapes}
\usetikzlibrary{arrows}
\usetikzlibrary{fit}
\usetikzlibrary{calc}
\usetikzlibrary{decorations.pathreplacing}

\tikzstyle{startstop} = [rectangle, rounded corners, minimum width=3cm, minimum height=1cm,text centered, draw=black, fill=red!30]
\tikzstyle{io} = [trapezium, trapezium left angle=70, trapezium right angle=110, minimum width=2.5cm, text width=2.5cm, minimum height=1cm, text centered, draw=black, fill=blue!30]
\tikzstyle{process} = [rectangle, minimum width=2.5cm, minimum height=1cm, text width=2.5cm, text centered, draw=black, fill=orange!30]
\tikzstyle{decision} = [diamond, aspect=3, minimum width=3cm, minimum height=1cm, text centered, draw=black, fill=green!30]
\tikzstyle{arrow} = [thick,->,>=stealth]
\tikzstyle{darrow} = [thick,<->,>=stealth]
\tikzstyle{entity_own} = [rectangle, minimum width=2.5cm, minimum height=1cm, text width=2.5cm, text centered, draw=black, fill=orange!30]
\tikzstyle{entity_foreign} = [rectangle, minimum width=2.5cm, minimum height=1cm, text width=2.5cm, text centered, draw=black, fill=gray!10]
\tikzstyle{functionality} = [rectangle, minimum width=2.5cm, minimum height=1cm, text width=2.5cm, text centered, draw=black, fill=white]
\tikzstyle{dataset} = [rectangle, minimum width=2.5cm, minimum height=1cm, text width=2.5cm, text centered, draw=black, fill=blue!10]
\tikzstyle{circlenode} = [circle, inner sep=0.05cm, draw=black, fill=gray!10, text centered]
\tikzstyle{ordernode} = [circle, thick, inner sep=0.02cm, draw=black, text centered]

\newtcolorbox{takeawaybox}{
  breakable,
  colback=gray!5,
  colframe=black!25,
  boxrule=0.5pt,
  sharp corners,
  left=8pt,right=8pt,top=6pt,bottom=6pt,
  title=Takeaways,
  fonttitle=\bfseries,
  before upper={\setlength{\leftmargini}{1em}},
}

\newcommand{\votingSolPerEpoch}{\qty{2.16}{SOL/epoch}}

\newcommand{\votingSolPerDay}{\qty{1.08}{SOL/day}}

%% file: acronyms.tex
\begin{acronym}[Derp]
    \acro{apy}[APY]{annual percentage yield}
    \acro{dpos}[DPoS]{Delegated Proof of Stake}
    \acro{ecdf}[ECDF]{Empirical Cumulative Distribution Function}
    \acro{lst}[LST]{Liquid Staking Token}
    \acro{pos}[PoS]{Proof of Stake}
    \acro{poh}[PoH]{Proof of History}
    \acro{posd}[PoSD]{Proof of Stake with Delegation}
    \acro{pow}[PoW]{Proof of Work}
    \acro{pp}[PP]{percentage point}
    \acro{sfdp}[SFDP]{Solana Foundation Delegation Program}
\end{acronym}

%% file: sections/introduction.tex
\section{Introduction}\label{sec:introduction}
The security of a blockchain is tied to who participates in the consensus and how much voting power each participant holds. Most consensus protocols rely on a threshold assumption.
The system remains secure as long as adversarial validators control less than a certain fraction of the total voting power,
typically one-third or one-half~\cite{nakamoto2008bitcoin,castro1999practical,kiayias2017ouroboros}.
Thus, the mechanism used to assign validator roles and distribute the voting power is crucial for security and resilience.
Despite over a decade of production use, the blockchain community has not converged on which consensus design best balances decentralization, security and scalability, and this trade-off remains an active subject of comparative study~\cite{andrews2025comparative,motepalli2025decentralization}.

The first widely adopted approach was \ac{pow}, which binds voting power to computational resources~\cite{nakamoto2008bitcoin,wood2014ethereum,vansaberhagen2013cryptonote}.
It has proved effective in open, adversarial environments. However, the competition for computational power makes \ac{pow} expensive in terms of energy consumption and hardware requirements.
This motivated \ac{pos} where voting power is bound to the amount of funds locked or staked in the protocol~\cite{kiayias2017ouroboros,buterin2019casper,kwon2014tendermint}.
This shift has enabled a new generation of blockchain systems that are significantly more resource-efficient. 
However, those who possess capital are not necessarily those who have the operational capacity, technical infrastructure or willingness to run validator nodes.

Thus, \ac{dpos} decoupled capital provision from direct participation in consensus by introducing two distinct roles~\cite{larimer2017dpos,eosnetwork,tronnetwork}:
\emph{Validators} operate the infrastructure and participate in consensus.
\emph{Delegators} then contribute stake without running validator infrastructure themselves, assigning their funds to validators of their choice.
However, classical \ac{dpos} systems (\eg EOS) restrict validator participation to a small, fixed committee elected by the community~\cite{larimer2017dpos,eosnetwork,tronnetwork,bitshares}, making it slow to adapt to changing network value. A newer generation of systems, which we call \ac{posd}, removes this bottleneck.
Any node can become a validator, with influence proportional to delegated stake. Solana, Cardano and Sui are among the largest deployments of this model.\footnote{We use \ac{posd} throughout to refer to this open-participation model, reserving \ac{dpos} for the classical, fixed-committee design it succeeds.}
In 2026, \ac{posd} secures 32 major blockchains with a total of $>$\$120B in native-token market cap~\cite{coingecko_pos,cryptoslate_dpos}.

\para{Contributions}
We contribute to the discussion about the optimal voting power allocation methods by investigating \ac{posd} using Solana as a case study.
Solana is, by far, the largest \ac{posd} blockchain and implements a canonical delegation mechanism, making it a perfect lens for studying the protocol dynamics. 
Existing theory~\cite{jensen1976theory,michels1911} indicates that \ac{posd} can only deliver an independent and sustainable financial system if it satisfies the following properties:

\para{P1: Power decentralization}\label{prop:decentralization} \ac{posd} must avoid excessive concentration of power among both validators and delegators. If a small number of validators collectively control a blocking or majority stake, the system becomes fragile.
Software failures, attacks or collusion affecting only a small handful of validators could disrupt consensus or compromise security~\cite{pan2021,kanti2026}.

\para{P2: Economic incentives}\label{prop:incentives} \ac{posd} must provide sustainable incentives for validators and delegators. Running a validator entails operational costs. These costs must be compensated for to make participation sustainable.
Delegators commit capital that is locked in the protocol and exposed to opportunity costs.
Thus, they also require sufficient rewards to justify participation.

\para{P3: Quality-based delegation}\label{prop:delegation} Delegation should track validator quality and reliability. If delegation is driven primarily by commission, delegators have little incentive to reward reliable operation, and the resulting stake allocation may fail to reinforce secure operation.

First, we investigate power centralization (P1) over time~(\cref{sec:centralization}).
Only 1.4--2.6\% of the largest validators jointly control a blocking (33\%) share of stake.
In several epochs,
validators from a single country control a blocking share of stake.
We also document a 44.5\% decline in the number of active validators over the observation period.
Concentration is even more extreme among delegators, despite their growing number. Just 0.003\% of the wealthiest delegators jointly hold a blocking share of total stake.
We also show a central role of institutional delegators
(\eg staking pools or delegation programs).

Second, we assess the economic incentives and sustainability (P2) of protocol participation~(\cref{sec:economics}).
We observe up to 45\% of validators operating at a loss, with the unprofitability being concentrated among validators that subsequently exit.
We also show how the total stake and the commission rates drive validators' profits. 
Delegator returns decline (average APY: 3.9\%$\to$3.5\% between September 2025 and March 2026), as validators retain a growing share of total rewards (30\%$\to$34\%).

Third, we examine whether delegators reward validator quality or simply chase the lowest price (P3),
by measuring what behaviors they respond to when allocating stake~(\cref{sec:delegator_behavior}).
Delegation program affiliation and commission changes are the strongest predictors of growth.
Validators approved by Solana's delegation program are more than twice as likely to attract additional stake and delegators, while a commission cut can triple the odds of subsequent growth.
On the other hand, reliability signals such as uptime show weak association with growth.

Finally, we discuss what these results imply for the design of \ac{posd} systems.
The structural problems we observe follow from the design of \ac{posd}.
We therefore argue that remedies should target the incentive structure and outline protocol-level changes.

%% file: sections/background.tex
\section{Background}
We provide background on Solana's implementation of \ac{posd}.
Other \ac{posd} systems (\eg Sui, Cosmos, Cardano) share its components.
We discuss how our findings generalize to them in \Cref{sec:discussion}.
In Solana, validator participation is permissionless.
Any node can become a validator, and its voting power scales with the stake delegated to it.
Time is divided into epochs, each consisting of \num{432000} slots.
Slots are scheduled to last $\approx \qty{400}{\milli\second}$,
resulting in $\approx$2-day epochs.
Validators perform two primary tasks: block production and voting.

\para{Block Production}
For every slot, a single validator is deterministically selected as the leader.
The leader builds and propagates a block containing user transactions. 
Leader selection is proportional to the validator's activated stake relative to total network stake.
Thus, validators with greater delegated stake are scheduled more frequently.
A leader receives user-paid fees for every transaction included in its block.
These block production rewards are not shared with delegators.

\para{Voting}
All validators continuously vote on newly produced blocks.
Each vote incurs a transaction fee of 0.000005\,SOL, totaling $\votingSolPerEpoch$.
At the end of each epoch, a fixed amount of newly minted SOL is allocated as voting rewards.
Each validator earns 1 epoch credit per successful vote on a finalized block.
Its share of the epoch's voting rewards is proportional to its stake and its accumulated epoch credits relative to other validators.
Each validator defines a commission rate representing the fraction of voting rewards retained by themselves.
The remainder is distributed to delegators.
Persistent failure to vote in time results in a validator being classified as delinquent. %
However, delinquency does not result in additional penalties beyond the loss of rewards for missed votes.

\para{MEV Rewards}
Validators can also earn maximal extractable value (MEV), \ie revenue from choosing which transactions enter a block and in what order.
On Solana, Jito runs out-of-protocol blockspace auctions through which validators earn MEV~\cite{helius_solana_mev}.
Searchers submit bundles of transactions, along with a tip bidding for guaranteed, ordered inclusion.
Jito retains a share of the tip before forwarding the remainder to the validator.
Validators define a separate MEV commission rate,
distinct from the voting commission,
that determines how the remaining tip is split with delegators.

\para{Delegation Program}
The \ac{sfdp} is a Solana Foundation initiative that supports new and eligible validators by covering their voting costs and matching a share of their delegated stake for 1 year, conditional on meeting performance requirements.%

\para{Delegator Rewards}
Delegators delegate their stake, held in a stake account, either directly to validators or to a stake pool.
The pool acts as an intermediary that abstracts validator selection and automatically rebalances stake across validators according to a defined strategy.
In both cases, the delegators receive the post-commission remainder of the validator's voting and MEV rewards proportional to their delegated stake.
Thus, delegators earn an \ac{apy} that depends only on the validator's commission and performance and is independent of the validator's total stake.

%% file: sections/methodology.tex
\section{Datasets}\label{sec:methodology}
We study Solana epochs 709--940, spanning 9 December 2024 to 15 March 2026.
We collect our on-chain data from Flipside~\cite{flipsidecrypto} and supplement with additional sources stated below.

\para{Validators}
We collect epoch-specific validator snapshots containing current stake, delegations, commission, epoch credits, delinquency, latitude, longitude and data-center identifiers, covering \num{2060} unique validators.
We define an active validator as a validator with an observed validator-epoch record in the relevant epoch.
We retrieve validators' \ac{sfdp} participation data from the \ac{sfdp} API~\cite{solanafoundationapi}.

\para{Delegators}
We collect epoch-level delegator (stake-account) snapshots
containing the validator assignment, delegated stake and staking authority associated with each stake account~\cite{flipsidecrypto}.
We identify stake-pools and liquid-staking accounts using stake-pool metadata from Solana Compass~\cite{solanacompass}.

\para{Delegation Relationships}
Staking actions contain delegation, deactivation and withdrawal operations.
We use these together with stake-account snapshots, linked via vote-account identifiers where required, to identify delegation relationships.

\para{Rewards and Prices}
Reward data contain validator and delegator rewards, including voting (inflation) rewards, block/transaction fees, priority fees and MEV rewards from Jito.
We use daily SOL prices to express rewards, revenue and operating costs in both SOL and USD.

%% file: sections/centralization.tex
\section{Power Centralization}
\label{sec:centralization}

We start by exploring the scale of power consolidation in Solana from the perspective of both validators and delegators (P1).
It is critical that
\ac{posd} avoids excessive concentration
to keep the system from becoming fragile~\cite{pan2021,kanti2026}.

\subsection{Validators}
We first examine the validator population and the distribution of stake across them.
Validators drive consensus, and delegated stake determines their voting power.
The number of validators and the concentration of stake therefore provide complementary measures of how widely consensus power is distributed.

The network maintains a large ($\approx 1k)$ active validator set.
This is substantially larger than the active validator sets of many other \ac{posd} systems (\eg Sui: 127~\cite{suiscan_sui_validators}, Cosmos: 180~\cite{cosmos_hub_faq} and Tezos: 192 validators~\cite{tezos_systems_bakers})
and is enabled by a high throughput and thus frequent user fees.
However, the number of validators declines by 44.5\% (1,397 $\rightarrow$ 775) over our observation period.
This can directly impact security, as a shrinking validator set reduces the number of malicious validators needed to corrupt the system (which we examine in \Cref{sec:economics}).

\begin{figure*}[htbp]
    \centering
    \begin{minipage}[t]{0.49\textwidth}
        \centering
        \includegraphics[width=\linewidth]{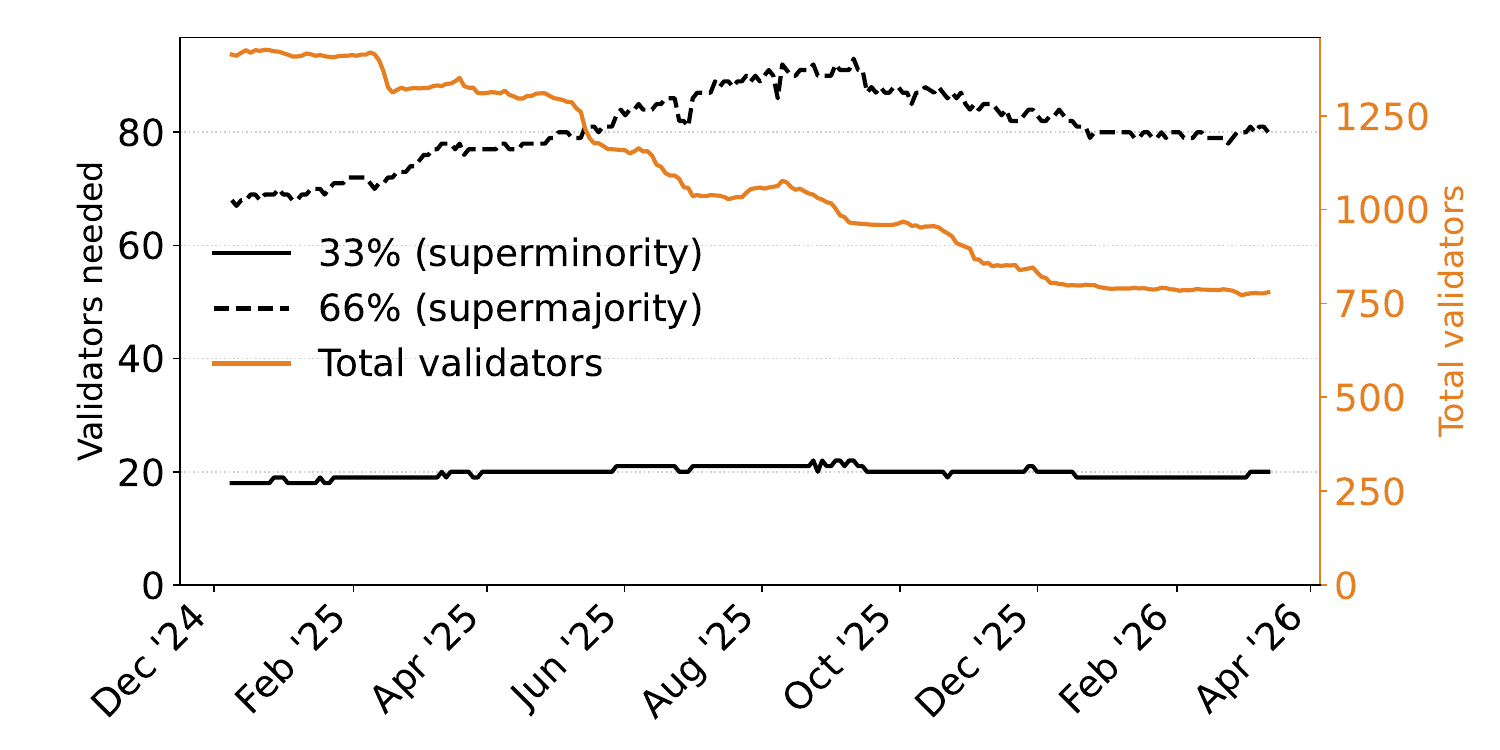}
        \captionof{figure}{Validator concentration.}
        \label{fig:centralization_validators}
    \end{minipage}
    \begin{minipage}[t]{0.49\textwidth}
        \centering
        \includegraphics[width=\linewidth]{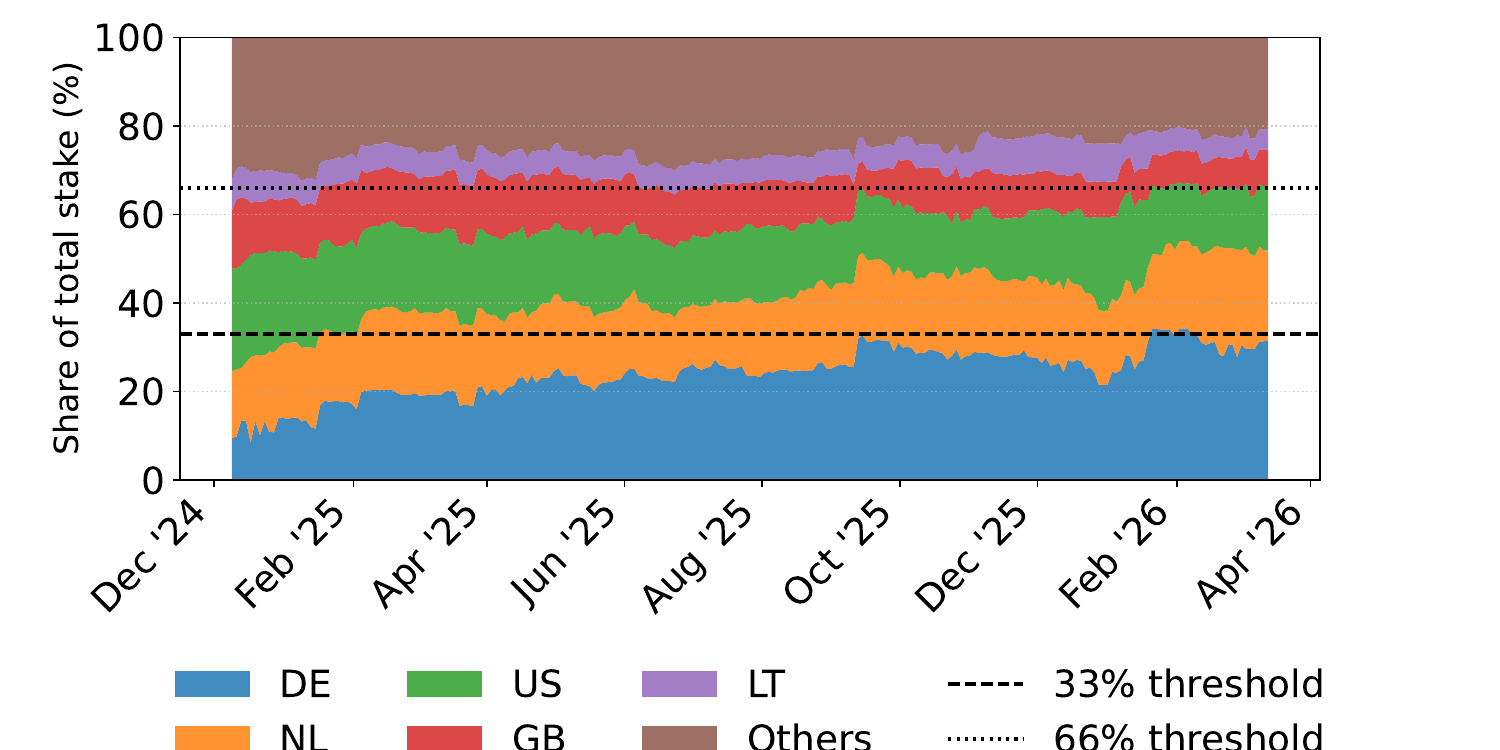}
        \captionof{figure}{Geographic stake distribution.}
        \label{fig:validators_geographic_distribution_over_time}
    \end{minipage}
\end{figure*}

\para{Consensus Thresholds}
\Cref{fig:centralization_validators} shows the minimum number of validators needed to control 33\% and 66\% of the voting power over time.
Both thresholds are important. 
An adversary controlling $>$33\% (superminority) can deny system liveness and finalize conflicting blocks under favorable network conditions.
At $>$66\% (supermajority), the adversary unilaterally controls the entire consensus.
We discover a high degree of centralization.
Across our measurement period, only 1.4--2.6\% (superminority) and 4.9--10.3\% (supermajority) of the largest validators by weight are capable of jointly blocking or hijacking consensus. 
Surprisingly, despite the earlier observed decline in the number of validators, the absolute number needed to control both thresholds remains stable at $\approx 20$ and $\approx80$ validators.
This indicates that those leaving the network are within the long tail of low-stake validators, while the richest set remains stable.

\para{Geography}
The same centralization risks arise from concentration
of validators within the jurisdiction of a small set of countries.
\Cref{fig:validators_geographic_distribution_over_time} plots the total stake of validators located in specific countries over time.
We again observe high and increasing centralization.
Validators located in Germany and the Netherlands already control >33\% of stake, while Germany alone reaches this threshold on specific dates (\eg January 2026).
The top 5 countries (\ie Germany, the Netherlands, US, the UK and Lithuania) reach the 66\% threshold and steadily grow their combined share
(70\% $\rightarrow$ 80\%).
This growth is driven mainly by Germany (10\% $\rightarrow$ 30\%) over our measurement period,
while the shares of the remaining countries remain comparatively stable.
Our investigation suggests that this is driven mostly by validator migration to Frankfurt-based data centers.%
Frankfurt hosts DE-CIX, one of the world's largest Internet exchange points, giving validators located there a lower latency to the rest of the network~\cite{cryptobriefing_de_dominance,cointribune_glassnode_europe}.
Lower latency to other validators 
improves leader-slot performance and rewards, making German validators more attractive to delegators independent of any cost differential.
Geographic concentration therefore creates correlated operational and regulatory risk because failures affecting a small number of countries can remove a substantial share of voting power.

\subsection{Delegators}

\begin{figure*}[t]
\centering
\begin{minipage}[t]{0.48\textwidth}
    \centering
    \includegraphics[
        width=\linewidth
    ]{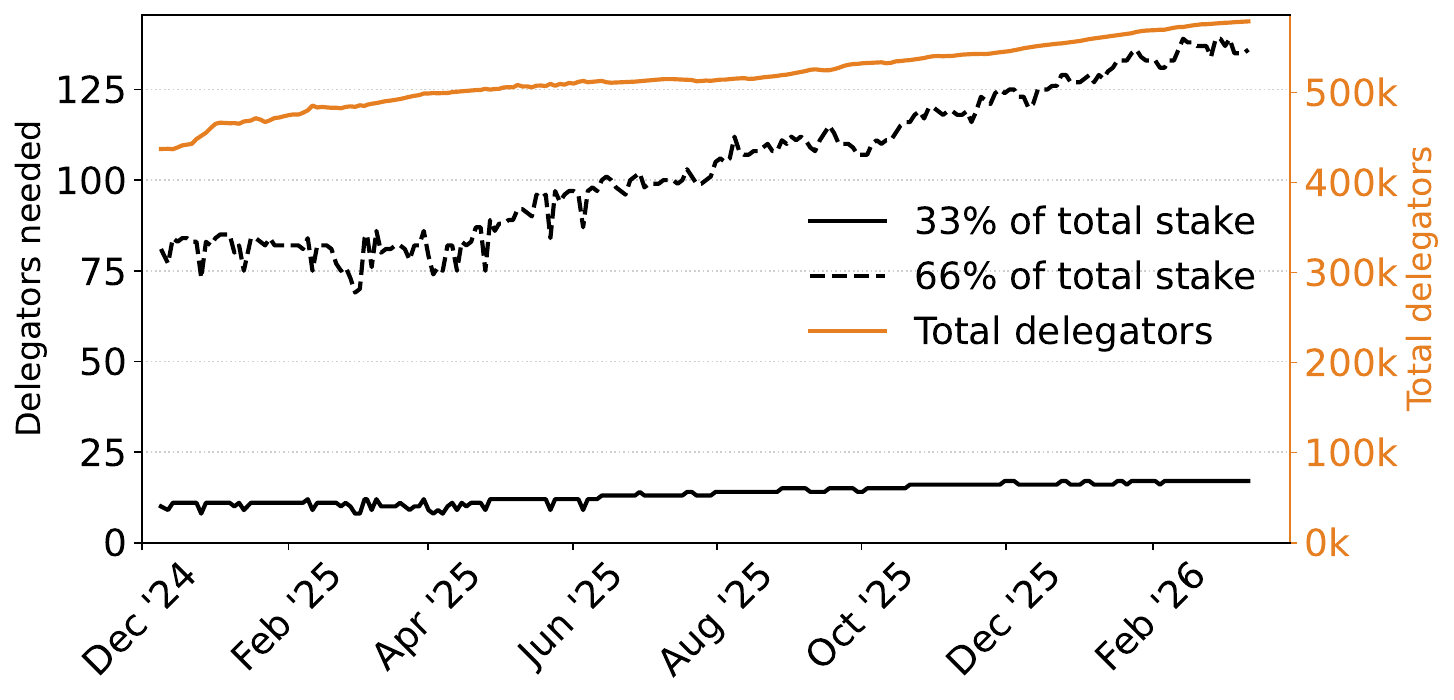}
    \caption{Delegator concentration.}
    \label{fig:delegator_concentration}
\end{minipage}
\hfill
\begin{minipage}[t]{0.48\textwidth}
    \centering
    \includegraphics[
        width=\linewidth
    ]{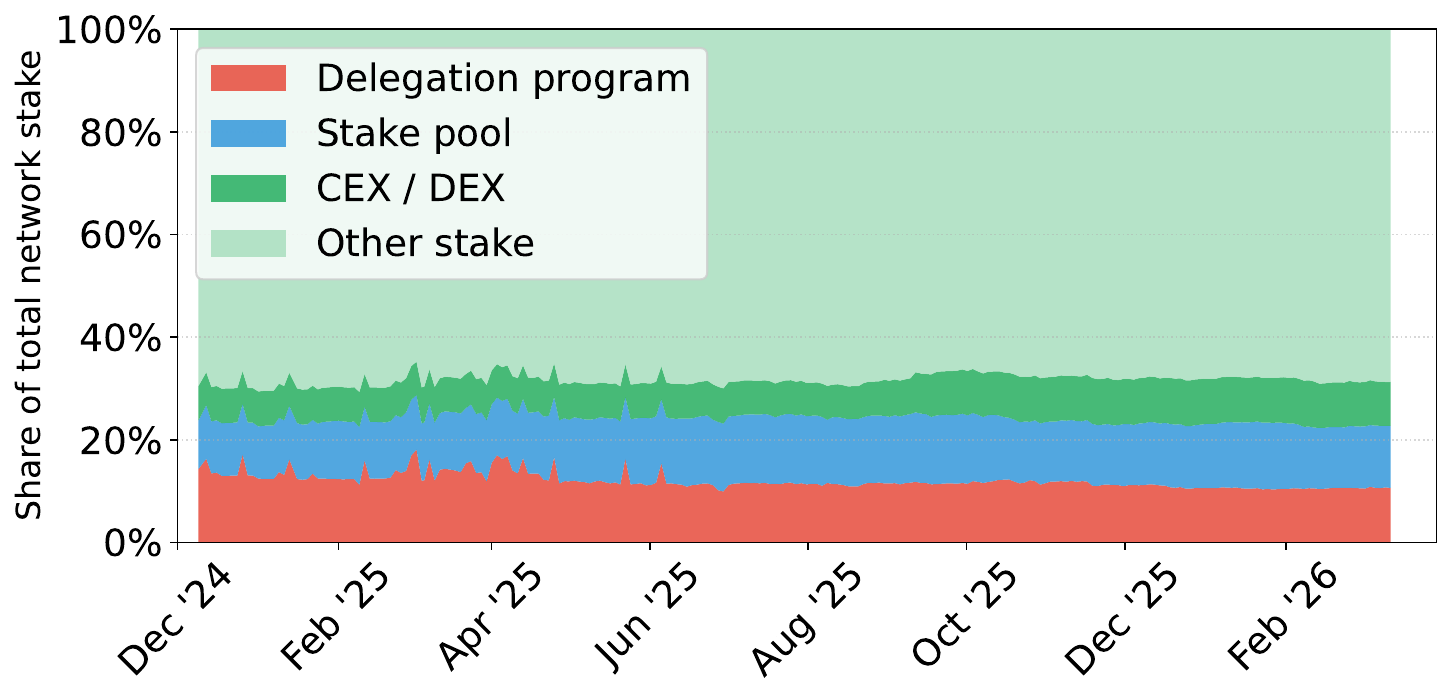}
    \caption{Stake across delegator types.}
    \label{fig:delegator_stake_distribution}
\end{minipage}
\end{figure*}

We next examine concentration within the delegator population.
Delegators give voting power to validators and thus directly shape the system power distribution.
We therefore aim to explore whether the delegators themselves (and their stake) remain centralized.
\Cref{fig:delegator_concentration} presents the number of delegators in the system over time (orange line).
Unlike validators, the total number of delegators grows by $>30\%$ (450k $\rightarrow \approx$600k).
This increase strongly outpaces the 9.5\% growth in total staked SOL.
Thus, the average stake held per delegator is falling, suggesting that the delegator population is in fact becoming more decentralized.

\para{Consensus Thresholds}
To confirm this, \Cref{fig:delegator_concentration} also inspects the number of delegators needed to reach 33\% and 66\% stake thresholds.
Despite new delegators joining and the average stake held per delegator falling, these stay remarkably low.
Only 10--17 delegators (0.003\%) are required to jointly hold 33\% of total stake, and 80--135 (0.02\%) to hold 66\%.
Delegator-side concentration is therefore far more severe than for validators.
We count delegators by staking authority, so a single account may be an intermediary (\eg a delegation program, stake pool or exchange) allocating the pooled funds of many owners.
These numbers thus measure control over stake allocation rather than ownership.
Note, this is a lower bound on the centralization degree as a single entity can control multiple delegator accounts. For instance, several accounts hold amounts close to exactly 5M SOL, indicating a potentially intentional split.
Such distribution is highly problematic as delegators act as kingmakers: their allocation determines which validators accumulate stake and voting power.
Thus, high concentration enables a small group of delegators to control the entire system.

\para{Delegator type}
Delegators can route their stake to validators through several distinct pathways, determining who actually controls the allocation decision.
Direct delegation leaves that decision with the original capital owner, whereas institutional delegation transfers it to an intermediary that allocates the pooled stake of many delegators at once.
The latter therefore gives increased tacit power to such institutions.
We thus investigate how the delegator stake is split across 4 categories:
\begin{enumerate*}
    \item individual delegators who stake directly,
    \item centralized and decentralized exchanges (CEX/DEX) that stake customer deposits on their behalf,
    \item stake pools and liquid staking protocols and
    \item delegation programs (\eg \ac{sfdp}).
\end{enumerate*}
\Cref{fig:delegator_stake_distribution} plots the total stake routed through each of these 4 categories.
The 3 institutional categories jointly control $>$33\% of total stake.\footnote{This is a lower bound on institutional control, as we cannot attribute all stake held in the ``other'' category.}
Delegation programs (17\%) constitute the largest institutional category,
followed by stake pools (10\%)
and CEX/DEX custodial delegation (8\%).
CEX/DEX and stake pools offer a compelling alternative for users. They automate optimal validator selection and maximize profit, but remove fund custody.
On the other hand, delegation programs aim to reduce validator-side centralization.
The Solana Foundation's program (5.7\%) subsidizes the economic viability of small and new validators,
DoubleZero (3.1\%) redirects stake toward validators in underrepresented regions (\eg São Paulo, Singapore)~\cite{doublezero_dzdp,doublezero_phase2},
and Firedancer (1.9\%) rewards validators for adopting an alternative client implementation~\cite{firedancer_delegation_program}.
However, while fighting for validator-side decentralization,
these programs concentrate delegation institutional power.

\begin{takeawaybox}
    Across validators and delegators, we observe high levels of centralization. Changes in raw participant counts do not translate into broader control, which remains concentrated within a stable set of the richest participants.
    We identify distinct centralization forces on both the validator side (\ie migration toward optimal network locations) and the delegator side (\ie institutional influence).
    This concentration makes the network's safety and liveness guarantees dependent on a small set of actors.
\end{takeawaybox}

%% file: sections/economics.tex
\section{Economics}
\label{sec:economics}
For \ac{posd} to be successful, it is necessary to provide sustainable incentives for validators and delegators.
At a minimum, for \ac{posd} to work, the validator operator costs must be covered, and delegators must accept the opportunity cost of committing capital.
We therefore next explore the economic incentives and assess whether validator and delegator participation is economically sustainable.

\subsection{Validators}

\label{sec:economics_validators}
We first assess validator sustainability by comparing the revenue generated by operating a validator with its estimated voting and infrastructure costs. We describe the full cost model in \Cref{app:cost_estimation}.

\para{Profitability}
\Cref{sec:centralization} showed a 44.5\% drop in the number of validators.
We classify a validator as a leaver if they leave the network for at least 5 consecutive epochs and do not reappear before the end of our observation period.
For each epoch, we mark a validator as profitable if their total income exceeds their total cost. We then average this profitability indicator over the observation period.
We hypothesize that those who leave may do so because they become unprofitable.
Indeed, 70.62\% of stayers are profitable on average over the observation period, while 78\% of leavers are unprofitable at the time they leave.
This suggests an economic cause behind the centralization result in \Cref{sec:centralization} and explains why the blocking-threshold set remains stable even as the overall population nearly halves.
Interestingly, 70.68\%
of leavers had previously received \ac{sfdp} support and became unprofitable after being removed from the program.
It further shows the impact institutional delegators have on the network: their support can determine both a validator's profitability and its continued presence in the network.

Some of this exit is arguably efficient rather than concerning. Over the same period, the SOL/USD exchange rate fell by roughly 60\% ($\$221 \rightarrow \$88$ per SOL), so the USD-denominated value of the network fell sharply even as its SOL-denominated stake grew modestly (\Cref{sec:centralization}). A smaller, better-capitalized validator set is a natural response to a shrinking economic base securing it, not necessarily a reliability or centralization failure in itself.

\begin{figure*}[t]
    \centering
    \begin{subfigure}[t]{0.32\textwidth}
        \centering
        \includegraphics[
            width=\linewidth
        ]{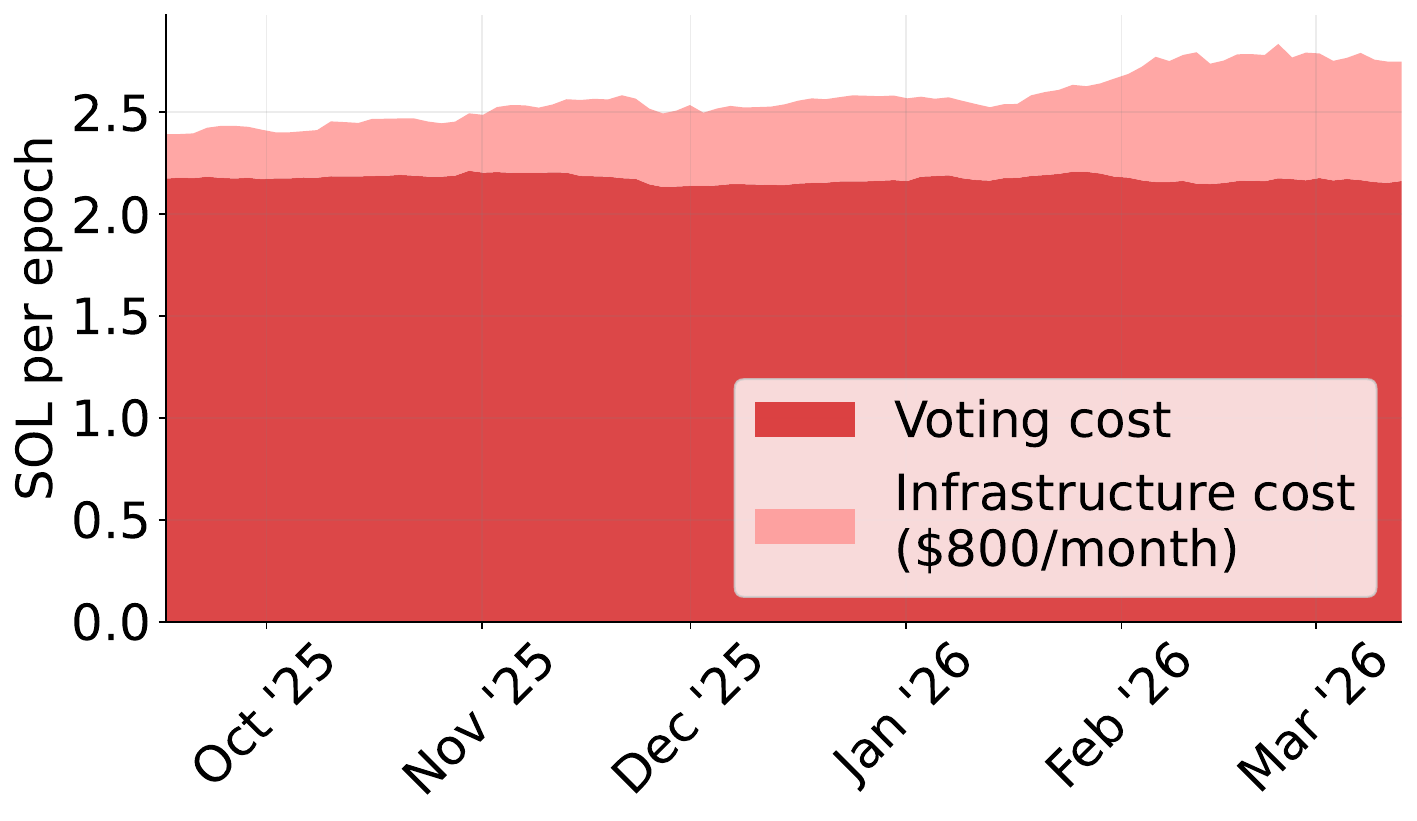}
        \caption{Validator costs.}
        \label{fig:solana_validator_voting_infrastructure_costs}
    \end{subfigure}
    \begin{subfigure}[t]{0.32\textwidth}
        \centering
        \includegraphics[
            width=\linewidth
        ]{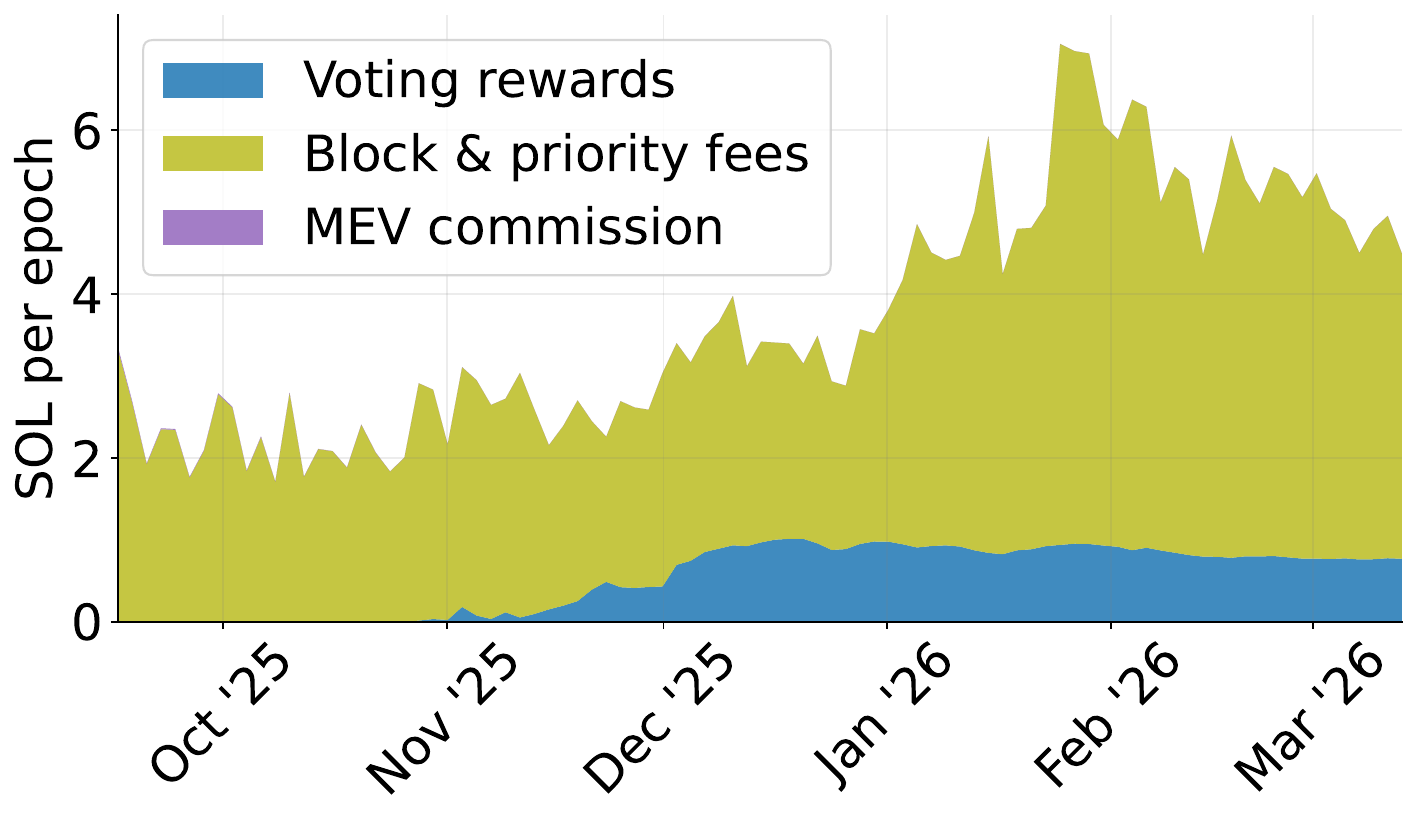}
        \caption{Median revenue.}
        \label{fig:solana_validator_median_revenue}
    \end{subfigure}
    \begin{subfigure}[t]{0.32\textwidth}
        \centering
        \includegraphics[
            width=\linewidth
        ]{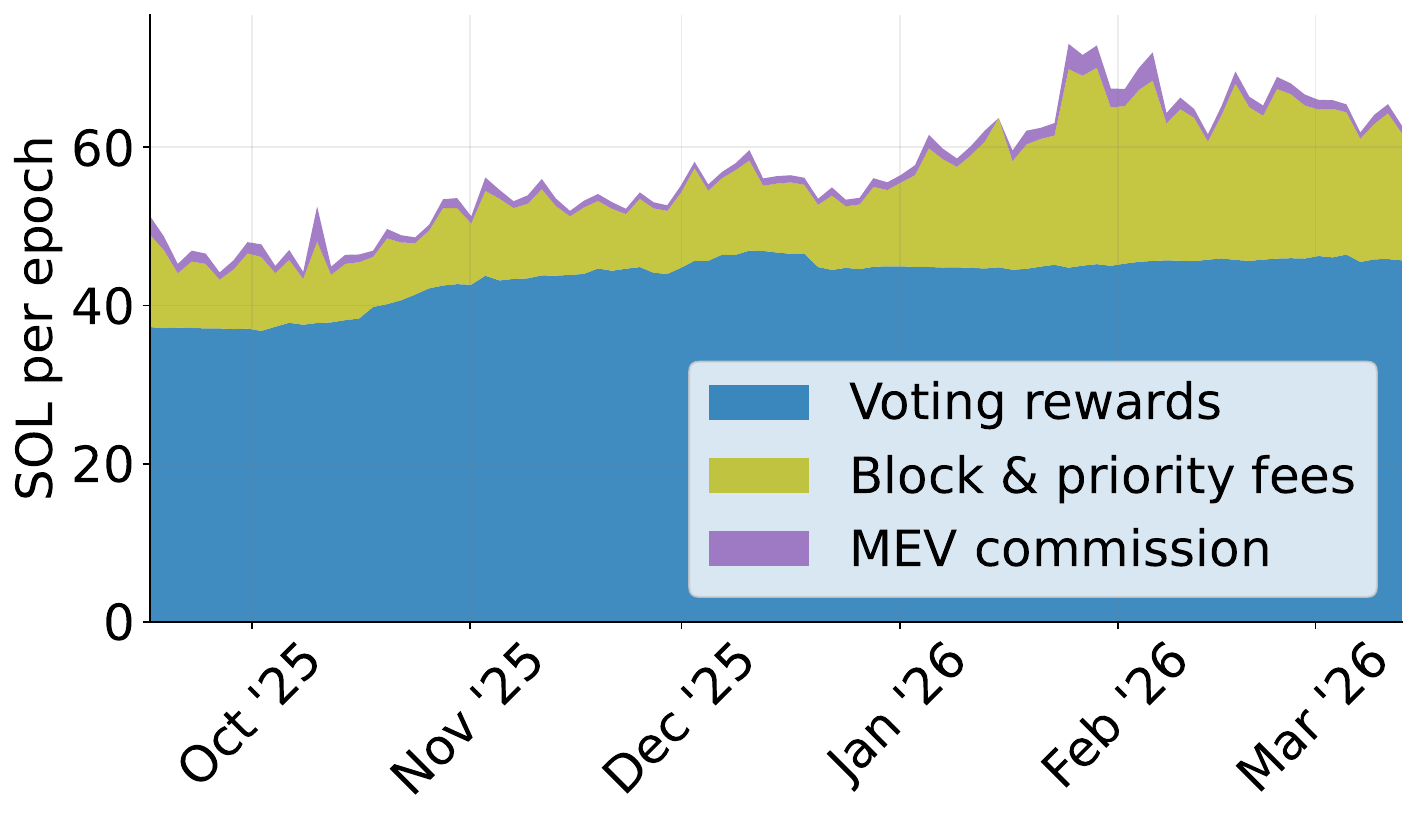}
        \caption{Mean revenue.}
        \label{fig:solana_validator_mean_revenue}
    \end{subfigure}
    \caption{Validator operating costs and revenue.
    }
    \label{fig:solana_validators_revenue_costs}
\end{figure*}

\para{Cost and Revenue}
We then ask why some validators are profitable while others are not.
\Cref{fig:solana_validators_revenue_costs} zooms in on the cost and revenue levels underlying this split. The costs are dominated by voting costs (80\%), but total costs increased ($2.4 \rightarrow $2.8\,SOL/epoch) due to the decreasing SOL/USD exchange rate.

Median revenue (\Cref{fig:solana_validator_median_revenue}) starts at only 2--3\,SOL/epoch, near the profitability margin, and grows to $\approx$5\,SOL/epoch as the least profitable validators leave the network.
Block fees and priority fees represent 85.49\% of the median income.
MEV commission is not an important source of validator revenue, representing only 0.04\% of median income. 
Mean revenue (\Cref{fig:solana_validator_mean_revenue}) is an order of magnitude above the median throughout (37--72 vs.\ 2--7\,SOL/epoch) and is instead dominated by voting rewards.
Costs, importantly, are largely independent of a validator's stake.
Every validator pays the same per-vote fee and requires broadly similar hardware, regardless of how much stake it commands.
Revenue, by contrast, scales directly with stake, since voting rewards are distributed in proportion to each validator's share of total staked SOL.
Thus, profit margins automatically improve with stake, giving large validators a structural advantage independent of operational efficiency.
This mirrors the stake concentration among the rich validators identified in \Cref{sec:centralization}.
The same mechanism that grants these validators outsized consensus power also grants them outsized economic returns, entrenching their position further. The sustainable-incentives property therefore holds only on aggregate, while small validators face a persistent viability gap.

\begin{figure*}[htbp]
\centering
\begin{minipage}[t]{0.48\textwidth}
    \centering
    \vspace{0pt}
    \parbox[t][3.4cm][t]{\linewidth}{%
        \centering
        \includegraphics[width=\linewidth]{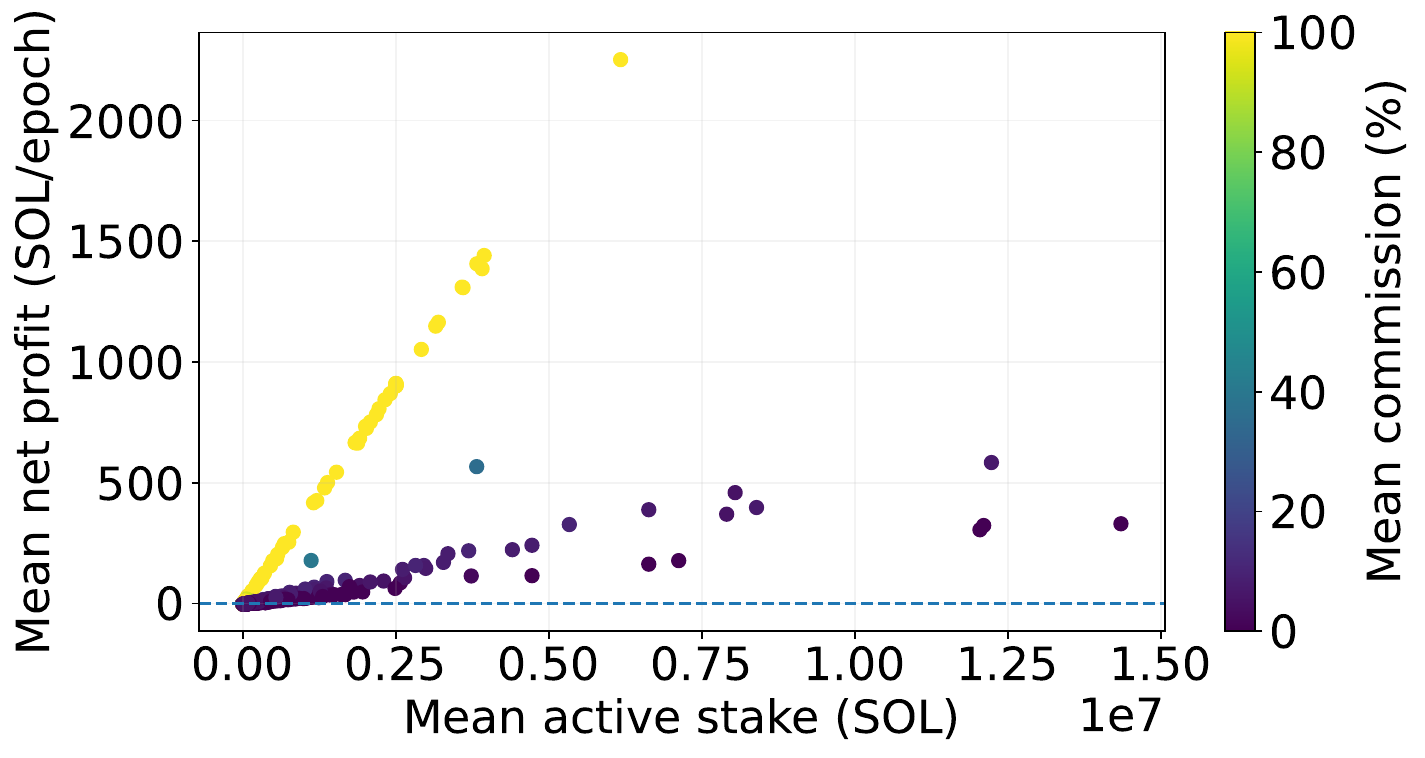}%
    }
    \captionof{figure}{Mean validator net profit.}
    \label{fig:solana_validators_profitability_with_commission}
\end{minipage}
\hfill
\begin{minipage}[t]{0.48\textwidth}
    \centering
    \vspace{0pt}
    \parbox[t][3.4cm][t]{\linewidth}{%
        \centering
        \includegraphics[width=\linewidth]{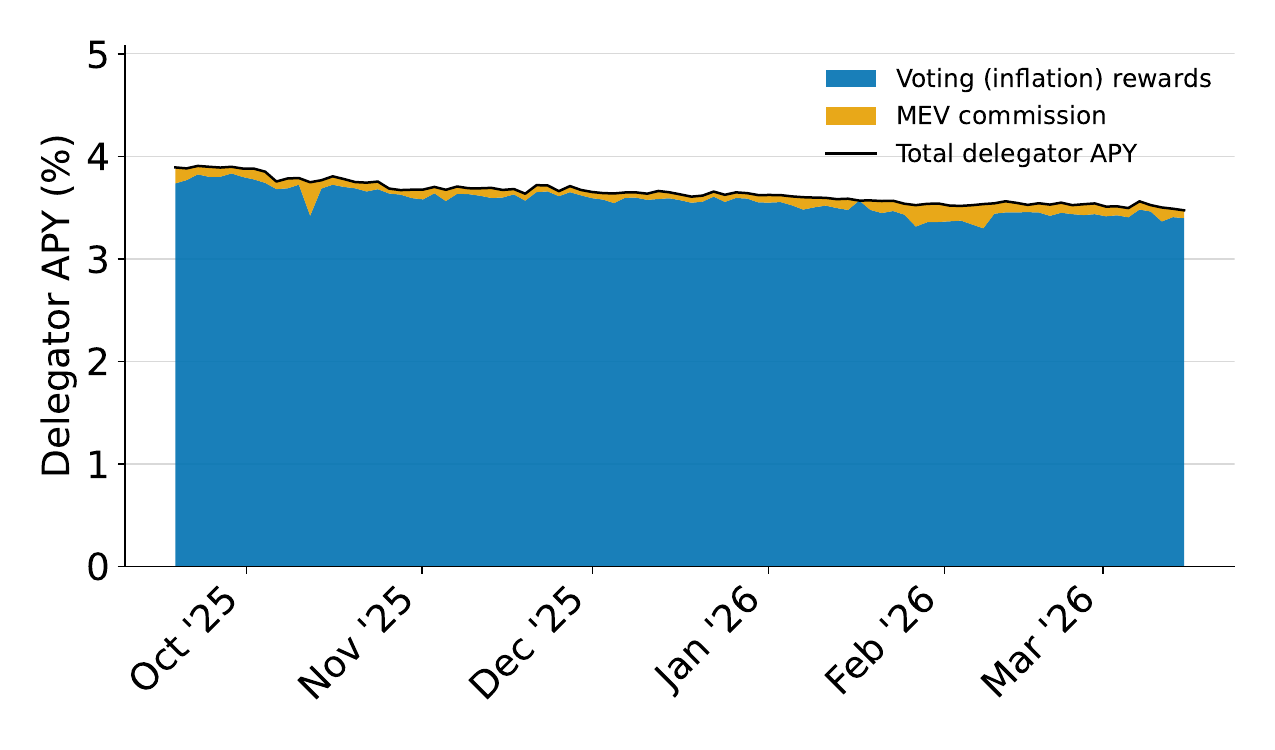}%
    }
    \captionof{figure}{Delegator APY.}
    \label{fig:delegator_apy_by_source}
\end{minipage}
\end{figure*}

\para{Profit Drivers}
\ac{posd} favors high-stake validators.
However, validator income is influenced by both stake (which affects block and voting income) and commission rate (which affects voting income alone).
We thus ask whether charging higher commission can compensate for having less delegated capital.
\Cref{fig:solana_validators_profitability_with_commission} plots mean net profit per epoch against mean active stake for each validator, colored by commission rate.
Stake does determine profitability.
All the net losers are grouped near the low-stake range, and average profit increases linearly with stake.
However, commission has an independent effect.
Validators charging 100\% commission trace a much steeper profit-stake line, earning roughly an order of magnitude more per SOL of stake than the 0--10\% group at comparable stake.
This also explains why mean revenue (\Cref{fig:solana_validator_mean_revenue}) is dominated by voting rewards.
A small number of high-commission validators retain a disproportionate share of their voting reward, pulling the population mean up.
However, this does not mean high commission is a good, long-term strategy. Charging more can drive delegators away over time. We further investigate this phenomenon in \Cref{sec:delegator_behavior}.

\subsection{Delegators}

We next assess if delegators are also incentivized to participate, based on whether returns to delegated capital are maintained.
We report this return per unit of delegated stake, annualized (\ac{apy}).

\begin{figure*}[htbp]
\centering
\begin{minipage}[t]{0.48\textwidth}
    \centering
    \includegraphics[
        width=\linewidth
    ]{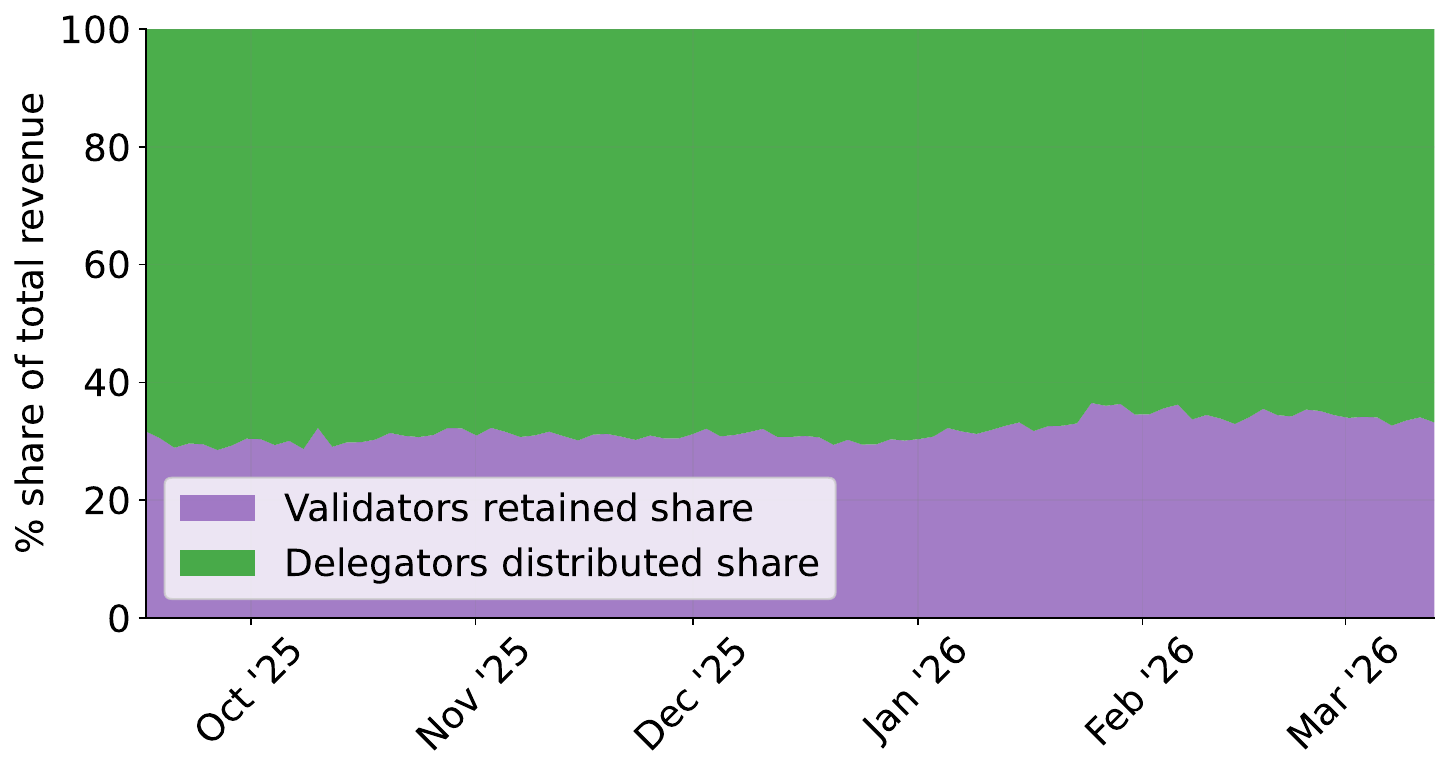}
    \captionof{figure}{Revenue share.}
    \label{fig:delegator_validator_revenue_share}
\end{minipage}
\hfill
\begin{minipage}[t]{0.48\textwidth}
    \centering
    \includegraphics[
        width=\linewidth
    ]{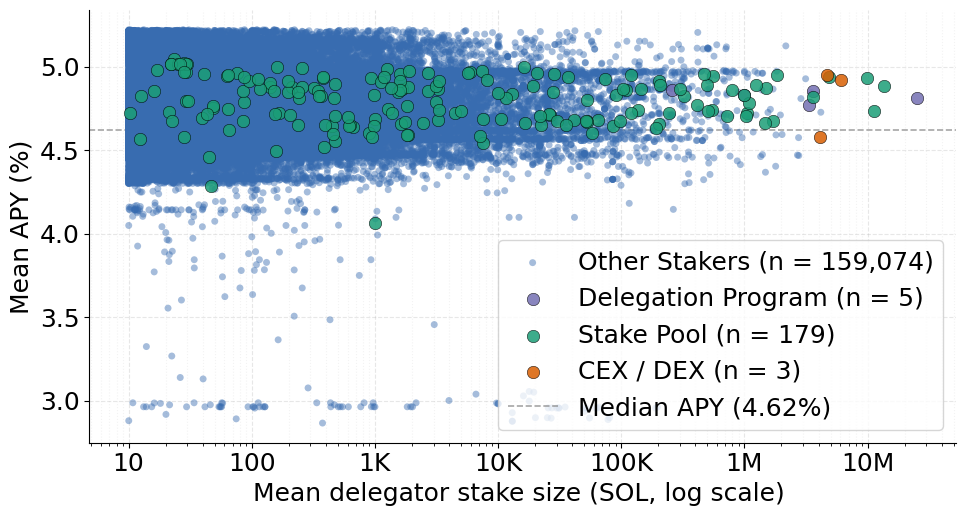}
    \captionof{figure}{Delegator APY distribution.}
    \label{fig:delegator_apy_distribution}
\end{minipage}
\end{figure*}

\para{APY}
\Cref{fig:delegator_apy_by_source} shows delegator APY over time.
The average revenue falls from September 2025 to March 2026 (3.9\% $\rightarrow$ 3.5\%), and comes almost entirely from voting rewards.
MEV commission is a small and shrinking share (4\% $\rightarrow$ 2\%).
Our investigation reveals 3 main causes.
First, Solana's disinflation (\cref{app:inflation_schedule})
schedule shrinks the total reward pool.
The inflation rate falls by 15\% a year toward a long-term target of 1.5\%~\cite{solana_inflation_schedule}.
Second, this shrinking reward pool is divided among a growing amount of delegated stake (\Cref{sec:centralization}), pushing the per-SOL yield down.
Third, validators keep a growing share of it for themselves by increasing the commission. \Cref{fig:delegator_validator_revenue_share} shows validators' share of total rewards rising (30\% $\rightarrow$ 34\%), with delegators' share falling (70\% $\rightarrow$ 66\%), consistent with the growing validator margin pressure documented in \Cref{sec:economics_validators}.

We investigate whether delegator type impacts APY. %
\Cref{fig:delegator_apy_distribution} plots mean APY against delegator stake size, split by delegation pathway. 
Institutional categories cluster around the median (4.62\%), regardless of stake size.
Individual delegators, by contrast, show far more spread.
Some earn above 5\%, while others earn only 3\% or less.
This is explained by the validator choice.
Institutional intermediaries appear to select established, medium-commission validators, delivering a stable, near-median return. Individual delegators either actively seek out low-commission validators, or stay with 100\% commission validators.
We observe 309 validators switching between 0\% and 100\% commission rates, potentially seeking to exploit delegators that do not pay sufficient attention. 
Notably, 2 validators do so particularly persistently, with 176 and 174 changes, respectively.
Thus, delegating through an institutional intermediary buys convenience and a predictable, near-median outcome, trading the upside of active selection for protection from its downside.

\begin{takeawaybox}
Validator's flat operating costs against stake-proportional revenue give large validators a structural advantage. A substantial share of smaller validators remain unprofitable and exit. On the delegator side, returns fall for 3 compounding reasons: a shrinking reward pool, more stake competing for it and validators retaining a growing margin.
Institutional intermediaries capture a safe, near-median yield while individual delegators bear the full spread of validator choice. \ac{posd}'s sustainable-incentives property (P2) therefore holds only on aggregate, and is weakest for the small validators and individual delegators with no structural advantage.
\end{takeawaybox}

%% file: sections/voting.tex
\section{Delegator Behavior and Validator Attractiveness}
\label{sec:delegator_behavior}
Stake allocation determines validator viability and the voting power distribution.
We thus investigate what drives delegation, and whether it tracks validator quality and reliability versus profit.
We separate validator features into static (persistent characteristics) and dynamic features (that change across epochs), and analyze their association with subsequent growth in total stake and in the number of unique delegators, using logistic regression. 
We provide feature definitions, model setup, full results and a discussion of causality in \Cref{app:delegator_behavior}.

\subsection{Static Features}

We first ask which persistent validator characteristics are associated with subsequent stake and delegator growth.
We call a validator's stake (or delegator count) \emph{grown} if it is higher at the validator's last observation than at its first.
Growth is thus a binary outcome, and we report how likely it is.
We start from a typical validator with average age and initial stake, no \ac{sfdp} or Jito participation and a location outside the major hubs.
The model gives this validator a 29\% chance of stake growth and a 16\% chance of delegator growth.
We then change one feature at a time, keeping the others fixed, and report the change in this probability in percentage points (pp).
A positive change means growth becomes more likely.

\para{Ecosystem Participation}
Ecosystem participation is the most visible certification signal available to a delegator who cannot otherwise verify validator quality.
\ac{sfdp} approval ($+$32\,pp) and Jito participation ($+$21\,pp) both raise the chance of stake growth.
The same holds for delegator growth ($+$25\,pp for \ac{sfdp} and $+$39\,pp for Jito). This strong association likely reflects 2 different mechanisms. 
Foundation support acts as a visible stamp of quality, attracting stake beyond the program's own direct contribution. Jito participation is plausibly still money-motivated.
Delegators are drawn to the promise of additional MEV-derived revenue, even though \Cref{sec:economics} shows this revenue is, in practice, only a small share of realized delegator APY. 
Delegators may therefore respond more to the prospect of extra yield than to its realized size.

\para{Age and Initial Stake}
Validator age and initial stake are negatively associated with subsequent growth.
Each additional standard deviation of age lowers the chance of stake growth ($-$7\,pp) and of delegator growth ($-$6\,pp).
Each additional standard deviation of initial stake has a smaller, similar effect on stake growth ($-$3\,pp) but little effect on delegator growth ($-$1\,pp).
Growth therefore concentrates among newer, smaller validators, suggesting a mild decentralizing force.
Established, high-stake validators are closer to a ceiling on further growth, while newer and smaller validators retain comparatively more room to attract additional delegation.
However, this describes the odds of \emph{any} growth from a validator's own baseline, not its amount. A small validator moving from near-zero to a modest stake still counts as growth, but does not reverse the persistent concentration.
The richest validators' structural revenue advantage (\Cref{sec:economics}) means their lead is not eroded by marginal growth elsewhere in the distribution.

\para{Infrastructure Location}
Infrastructure location is a signal that could plausibly proxy for either quality (network performance) or money (MEV capture).
For delegator growth, being hosted in one of the major hubs raises the chance of growth in Frankfurt ($+$9\,pp), Amsterdam ($+$7\,pp) and Tokyo ($+$15\,pp). These cities are also 3 of Jito's official Block Engine regions~\cite{jito_mainnet_addresses}.
Hosting there means proximity to MEV infrastructure.
Solana's 400\,ms block time leaves only a narrow window for the Block Engine to deliver bundles to the current leader, so validators farther from a Block Engine region risk bundles arriving too late to include, capturing less MEV during their own leader slots.

For stake growth, location has little effect.
None of the hubs meaningfully shift the baseline.
Delegator count is thus dominated by a large number of small, individual delegators, who are responsive to profit-maximizing signals such as proximity to MEV infrastructure (\Cref{sec:economics}).
At the same time, stake volume is dominated by a small number of large, often institutional delegators, who are more likely to follow formal criteria than a validator's data-center location.

\subsection{Dynamic Features}
The dynamic analysis asks whether changes in validator conditions (\ie commission and delinquency) are followed by changes in stake or delegator growth. We report the outcome after 1 and 20 epochs, as the short- and mid-run horizons.
Each analysis uses a different sample of validators (those with an event) and thus has its own baseline.

\para{Commission Changes}
We first analyze commission changes as an observable price signal.
Growth becomes more likely over longer horizons, so we compare each horizon with its own baseline (stake: 47\% after 1 epoch and 57\% after 20; delegators: 25\% and 53\%).
A commission increase lowers the chance of stake growth after 1 epoch ($-$20\,pp) and after 20 epochs ($-$19\,pp).
The drop in delegator growth is $-$14\,pp after 1 epoch and $-$23\,pp after 20.
A commission decrease raises the chance of stake growth after 1 epoch ($+$19\,pp) and after 20 epochs ($+$22\,pp).
For delegator growth, it has almost no effect after 1 epoch ($-$2\,pp) but a large one after 20 epochs ($+$24\,pp).

Stake thus reacts to a change in either direction within a single $\approx$2-day epoch.
This reflects active, profit-maximizing delegators (see \Cref{sec:economics}), including automated stake pools and liquid-staking protocols that rebalance algorithmically and react almost immediately.
Delegator counts move more slowly.
After an increase, delegators start leaving at once and keep leaving.
After a decrease, they do not respond at first.
Manual delegators plausibly discover the cheaper validator only over the following weeks, via staking dashboards and explorers.
Leaving a validator that became more expensive is thus quick, while finding one that became cheaper is slow.

\para{Delinquency}
We then analyze delinquency as an observable reliability signal.
At 1 epoch, the baseline chance of stake growth is 57\%, compared with 15\% for delegator growth.
For stake growth, a delinquency event followed by recovery raises the chance after 1 epoch ($+$7\,pp) and after 20 epochs ($+$16\,pp).
Persistent delinquency lowers it after 1 epoch ($-$13\,pp) and after 20 epochs ($-$6\,pp).
The effect on delegator growth is more pronounced.
Recovery lowers the chance after 1 epoch ($-$4\,pp) but raises it after 20 epochs ($+$5\,pp).
Persistent delinquency lowers the chance after 1 epoch ($-$10\,pp), with the drop reaching $-$24\,pp after 5 epochs and remaining at $-$12\,pp after 20 epochs.

However, the 20-epoch cohort is a shrinking, survivorship-biased sample, restricted to validators that had a single delinquency event and then operated cleanly for the following 20 epochs.
This is a meaningful signal.
Delinquency events are otherwise rare among healthy validators: 81\% of validators in our sample never experience one at all, and the leaver population accounts for a disproportionate share of the rest (54.43\% of leavers experience delinquency, versus only 0.64\% of stayers).
Surviving 20 clean epochs after a rare event therefore demonstrates that the underlying issue was transient, and delegators appear to reward that demonstrated recovery.
Delegators thus pay attention to reliability, punishing an outage immediately but returning after durable recovery.

\begin{takeawaybox}
Commission cuts and Jito participation drive growth.
For quality, delinquency is punished immediately and forgiven only after a credible recovery.
SFDP participation functions as a kingmaking visibility signal.
\ac{posd}'s quality-based delegation property (P3) therefore holds only weakly.
Delegators reward price and institutional signaling far more consistently than operational reliability.
\end{takeawaybox}

%% file: sections/discussion.tex
\section{Discussion}
\label{sec:discussion}
Building on our results, we provide recommendations for \ac{posd} design.

\para{Sublinear Stake-Weighted Rewards}
\ac{posd} gives large validators an advantage that grows with every additional unit of delegation. We propose making validator rewards a sublinear function of a validator's own stake, so that the marginal reward per additional SOL declines as stake rises above the network median. This would preserve the legitimate signal that a large stake carries (\ie validator has attracted delegator trust and support) while damping the compounding effect.
It would let delegators trade off yield-chasing (favoring smaller validators) against safety (an established, high-stake operator), without preventing poorly-run validators from becoming unprofitable and leaving, to preserve the healthy exit dynamics.

Polkadot's Nominated Proof-of-Stake pursues the same goal differently.
Its election algorithm reassigns nominator support to maximize the stake backing the worst-backed elected validator.
The system has a hard oversubscription cutoff that makes a delegator backing an already-popular validator earn no reward on that stake~\cite{cevallos2021npos}.
In practice however, this has made direct nomination's effective minimum stake requirement high and volatile, pushing most small holders toward pooled ``Nomination Pools'' instead and potential concentration~\cite{polkadot_nomination_pools}.
Classical DPoS instead assigns each elected validator roughly equal block-production reward through round-robin scheduling.
However, this does not itself resolve concentration.
For instance, EOS's vote-weighted election has produced persistent ``voting gangs'' among large stakeholders, voter turnout below 5\% and proxy accounts that now control roughly 70\% of voting weight~\cite{liu2022understanding}.

\para{Delegation Strategies}
Delegator returns in \ac{posd} are similar overall, but attentive delegators who react quickly still gain an advantage over inattentive ones.
The institutional intermediaries can offer a stable, near-median return to delegators who prefer convenience over active selection.
However, institutional intermediaries already control a meaningful share of Solana's delegated stake (\Cref{sec:centralization}), and comparable concerns are documented on Ethereum~\cite{grandjean2023ethpos}.
We argue that this is a major risk only when institutional players use opaque or discretionary validator selection logic.

We recommend that pooled-staking selection logic be published, executed on-chain and react automatically to observable signals such as commission and delinquency.
Stakers should be able to choose among several competing published strategies.
Solana's Jito Steward already implements a close analog, computing validator eligibility and scores directly on-chain from 10 published, automatically-checked criteria (including exclusion of superminority validators) and executing delegation accordingly~\cite{jito_steward}. Marinade Finance publishes an equally precise scoring formula and lets stakers direct a share of stake by vote instead of the algorithm~\cite{marinade_delegation_strategy}.
However, its score is computed off-chain, which requires trust in the computation itself. Neither system, however, lets stakers choose among multiple competing published algorithms the way a traditional investor chooses between index-fund methodologies.

\para{MEV-Driven Geographic Concentration}
\Cref{sec:delegator_behavior} shows that Frankfurt, Amsterdam and Tokyo's role as Jito Block Engine regions already pulls both validators and delegators toward these locations.
This happens even though MEV currently represents only a small share of total revenue (\Cref{sec:economics}).
This pull is thus likely to strengthen.
MEV adoption is still rising, voting rewards are shrinking under Solana's disinflation schedule (\Cref{app:inflation_schedule}) and MEV-extraction strategies continue to mature.
All of that raises MEV's share of total reward over time.
This also creates a growing reliance on a single piece of external infrastructure as a point of centralization.

Ethereum faces a structurally similar problem and has pursued 2 distinct responses. The first accepts that MEV extraction naturally centralizes, and isolates that centralization from the validator set itself.
Proposer-builder separation (PBS) lets specialized, centralized ``builders'' construct blocks and compete for inclusion, while the decentralized set of validators retains the voting ~\cite{ethereum_pbs_roadmap,eip7732}. The second response instead limits MEV extraction.
Encrypted mempools, such as Shutter Network's threshold-encryption design, hide transaction content until after ordering is finalized, closing off the front-running and sandwich-attack vectors that exploit proximity to block-production infrastructure~\cite{shutter_network}.
More broadly, MEV is value extracted from ordinary users~\cite{daian2020flashboys}.
Recent theoretical work shows that no fee mechanism can be simultaneously welfare-maximizing for users and revenue-maximizing for MEV-extracting block producers~\cite{bahrani2024posmevworld}. This suggests that a network which limits MEV extraction could attract more legitimate usage and thus revenue through ordinary fees and rewards, even without MEV income.

\para{Generalizability}
Our recommendations rest on Solana, but other \ac{posd} systems (\eg Sui, Cosmos, Cardano) share its core components.
Validator entry is open, delegators choose whom to stake with, voting power follows stake and validators set a commission on the rewards they share.
The systems differ mainly in parameters that tune how strongly stake translates into power and reward.
Sui caps a validator's voting power at 10\%~\cite{sui_tokenomics}, Cosmos limits the active set to 180 validators~\cite{cosmos_hub_faq} and Cardano saturates pool rewards~\cite{cardano_staking_parameters}.
Such rules already work against the large-validator advantage we document, so we expect that advantage to be weaker there.
Commission-driven delegation should hold wherever validators set their own commission.
Some effects are specific to Solana.
The location pull, for example, stems from Jito's Block Engine regions and Solana's short slots.

%% file: sections/related_work.tex
\section{Related Work}
\label{sec:related_work}

Resource concentration is well documented in permissionless blockchains.
Beikverdi and Song~\cite{beikverdi2015} show that Bitcoin mining power aggregates in a few pools over time.
Gencer \etal~\cite{gencer2018} extend this to Ethereum and argue that decentralization is multidimensional and must be measured rather than assumed.
Wang \etal~\cite{9160462} show that pool incentives such as variance reduction and stable payouts reinforce this concentration, and Chen \etal~\cite{chen2020} propose Endex, a metric that treats decentralization as a measurable property.
Kwon \etal~\cite{10.1145/3318041.3355463} further argue that economic and communication constraints make full decentralization unattainable.

Work on \ac{dpos} reaches similar conclusions.
Jeong~\cite{jeong2020centralized} calls \ac{dpos} governance \enquote{centralized decentralization,} as voting-based delegation concentrates control among a few active participants.
Rebello \etal~\cite{9955597} show that quorum-based consensus trades off scalability, fault tolerance and decentralization.
Li and Palanisamy~\cite{10.1007/978-3-030-59638-5_2} find that \ac{dpos} is more efficient than \ac{pow} but more prone to validator concentration.
Karakostas \etal~\cite{karakostas2020account} formalize delegation to stake pools in \ac{pos} wallets and analyze its security implications.
The closest work is the data-driven study of Liu \etal~\cite{liu2022understanding}, who analyze how decentralization evolves in EOSIO.
However, their setting is classical \ac{dpos}, where voters elect a fixed-size committee of block producers.

%% file: appendix/appendix.tex
\section{Additional Details}

\subsection{Concentration Metrics}
\label{app:concentration_metrics}

\para{Stake Concentration}
For each epoch, we aggregate delegated stake by validator (or, for delegator-level analysis, by observed staking authority) and rank accounts by stake in descending order. The Nakamoto coefficient is the minimum number of accounts required to control at least one-third of total stake, $N = \min_k\left\{k \mid \sum_{i=1}^{k}s_i > 1/3\right\}$, where $s_i$ is the stake share of account $i$; we also calculate the corresponding coefficient at the two-thirds threshold. We additionally calculate the Gini coefficient over stake shares. These calculations include stake self-delegated by validators, stake delegated by external token holders and stake allocated through stake pools.

\para{Geographic Concentration}
For each epoch, country-level stake shares $s_i$ are calculated from the stake assigned to validators in country $i$, ordered by decreasing share, with $M$ the number of countries observed in the epoch. We measure geographic concentration using the geographic Gini coefficient, the geographic Herfindahl--Hirschman Index, $\mathrm{HHI}_g = \sum_{i=1}^{M}s_i^2$, and the geographic liveness coefficient~\cite{sen2026geographical}, $LC_g = \min\{k \mid \sum_{i=1}^{k}s_i > 1/3\}$.

To measure validator proximity to the geographic stake distribution, we calculate a fixed stake-weighted geographic center using validators with positive stake and non-missing coordinates in epoch 709, taken as the weighted mean of their latitude and longitude. For validator $v$, the distance to this center is calculated using the Haversine formula and used as a covariate in the validator-lifetime analysis. Data-center crowding is the number of validators that share the same data-center identifier.

\subsection{Validator Operating Cost Estimation}
\label{app:cost_estimation}

We estimate annual validator operating costs as the sum of voting and infrastructure costs, applying the same cost schedule to all validators as a common benchmark. Voting costs are calculated from the per-vote transaction fee and the expected number of votes per epoch, resulting in $\votingSolPerDay$. Based on Anza's hardware requirements,\footnote{\url{https://docs.anza.xyz/operations/requirements}} we calculate the cost of renting the infrastructure in the cloud and convert it to SOL using the mean hourly USD/SOL exchange rate within each epoch. Profitability is defined as total validator revenue minus these estimated operating costs. Because infrastructure requirements and costs may vary between operators, this measure is an estimate of revenue relative to a common operating-cost benchmark rather than realized accounting profit.

\subsection{Solana Inflation Schedule}
\label{app:inflation_schedule}
Solana distributes staking rewards each epoch by issuing new SOL according to a predefined inflation schedule. The initial inflation rate is set to 8\%. This rate decreases over time at a disinflation 15\% annual disinflation rate, reaching a long-term target of 1.5\%. The initial inflation rate refers to the starting annual rate when inflation is first activated and represents an upper bound that can only decline over time. The disinflation rate describes the speed at which this inflation rate is reduced each year. The long-term inflation rate is the stable equilibrium level that the system is expected to converge to over time.

\subsection{Solana Class of Validator}
\para{Centralized Exchanges}
Centralized exchanges provide a one-click staking service for off-chain exchange users. Examples include Kraken, Coinbase, Binance and Upbit. These entities are among the largest Solana validators by total stake.

\para{Insitution solution providers}
These companies specialize in staking services tailored to institutional clients. They support many blockchains to meet broader customer needs. Examples include Figment, Kiln and Twinstake.

\para{Institutional Staking Providers}
These entities specialize in staking services tailored to institutional clients. They typically support multiple blockchains to meet broader customer needs. Examples include Figment, Kiln and Twinstake.

\para{Private Validators}
Solana has plenty of private validators. Their stakes are self-delegated and likely controlled by the operating entities. These validators are usually characterized by 100\% commission rates and lack of public identity information on block explorers and dashboards.

\subsection{Comparison of Validators}
\label{app:validator_example}

Validators with superior performance can generate higher gross rewards even if they charge higher commission. Consider 2 validators:

\begin{itemize}
    \item Validator A: $c = 0$, $R_g = 6.5\%$
    \item Validator B: $c = 10\%$, $R_g = 7.5\%$
\end{itemize}

The corresponding net returns are:
\begin{align}
R_n^{A} &= 6.5\%, \\
R_n^{B} &= (1 - 0.10)\cdot 7.5\% = 6.75\%.
\end{align}

Despite higher commission, Validator B can deliver a higher net yield due to higher gross performance.

\para{Sources of APY Variation}
Observed \ac{apy} may vary due to short-term randomness in leader slot assignment, temporary network congestion affecting fee revenue, changes in total network stake and epoch-level reward fluctuations.

\begin{table}[h]
\centering
\caption{Pearson correlation coefficients between validator exit (IS\_LEAVER) and operational, geographic and performance variables.}
\label{tab:corr_exit}
\begin{tabular}{l r}
\hline
Feature & Correlation \\
\hline
IS\_LEAVER (baseline) & 1.000 \\
Delinquent status & 0.551 \\
Data center congestion level & 0.113 \\
Distance to cluster center (km) & -0.111 \\
Epoch credits & -0.650 \\
\hline
\end{tabular}
\end{table}

\begin{table}[h]
\centering
\caption{Validator performance by data center congestion level.}
\label{tab:dc_congestion}
\begin{tabular}{l r r r r}
\hline
DC bin & Mean epoch credits & Delinquency (\%) & Churn (\%) & N validators \\
\hline
Low (1--3 nodes)      & $4.23\times 10^6$ & 33.13 & 56.04 & 323 \\
Medium (4--10 nodes)  & $4.35\times 10^6$ & 32.49 & 56.22 & 434 \\
iCrowded (11+ nodes)   & $4.05\times 10^6$ & 34.75 & 65.49 & 1,298 \\
\hline
\end{tabular}
\end{table}

\begin{table}[h]
\centering
\caption{Performance and churn outcomes at the most densely populated data centers.}
\label{tab:dc_top}
\begin{tabular}{l r r r r}
\hline
Data center & Nodes & Mean credits & Delinquency (\%) & Churn (\%) \\
\hline
18450-US-Ogden & 84 & 3.10$\times 10^6$ & 45.24 & 86.90 \\
395201-DE-Fechenheim & 79 & 4.19$\times 10^6$ & 32.91 & 50.63 \\
20326-NL-Amsterdam & 75 & 4.35$\times 10^6$ & 44.00 & 49.33 \\
58061-NL-Amsterdam & 66 & 3.96$\times 10^6$ & 39.39 & 93.94 \\
16125-LT-Europe/Vilnius & 54 & 3.95$\times 10^6$ & 29.63 & 83.33 \\
47447-DE-Europe/Berlin & 43 & 4.48$\times 10^6$ & 23.26 & 81.40 \\
396356-DE-Frankfurt am Main & 41 & 3.72$\times 10^6$ & 34.15 & 60.98 \\
20326-DE-Frankfurt am Main & 40 & 5.42$\times 10^6$ & 12.50 & 30.00 \\
29066-DE-Frankfurt am Main & 39 & 3.97$\times 10^6$ & 30.77 & 82.05 \\
396356-US-New York & 36 & 3.40$\times 10^6$ & 61.11 & 75.00 \\
\hline
\end{tabular}
\end{table}

\subsection{Delegator Behavior}
\label{app:delegator_behavior}
This section supports \Cref{sec:delegator_behavior}.
It gives the feature definitions and model setup, the full regression results and a discussion of how to interpret them.

\para{Methodology: Static Features}
We consider the following static features:

\begin{itemize}
\item Initial stake: the validator's delegated stake at its first observed snapshot.
\item Validator age: the number of days between validator creation and the corresponding observation.
\item Ecosystem participation: indicators for participation in the \ac{sfdp},
including Approved or TestnetOnboarded, Rejected and Retired states, as well as an indicator for Jito participation. The SFDP indicators represent the mean status indicator across observed snapshots, while Jito participation is coded as 1 if it is observed at any snapshot. Jito participation is a separate, MEV-related indicator from SFDP status.
\item Infrastructure location: indicators for the 20 most frequent city groups derived from the data-center identifier. The modal city group is used for each validator, and all remaining locations, including unknown locations, are grouped into the reference category.
\end{itemize}

Before estimation, initial stake and validator age %
are standardized to mean zero and unit variance using a standard scaler. The intercept therefore represents a validator with average values of these features, no SFDP or Jito participation and the reference infrastructure-location category.

We create a single observation per validator, with 2 binary outcomes: whether its total stake and number of unique delegators were higher at its last observed snapshot than at its first. We regress each outcome on the features above using the logistic model introduced above. This allows us to ask whether a validator having a particular feature is more or less likely to experience subsequent growth. For each feature, we report the odds ratio ($\exp(\beta_j)$): a value above 1 indicates higher odds of growth associated with that feature (\eg holding SFDP approved status rather than not, or a 1-standard-deviation increase in a standardized continuous feature such as age), while a value below 1 indicates lower odds.

\para{Methodology: Dynamic Features}
Our second analysis examines events that change during validator operation, testing whether changes in commission and validator reliability are followed by changes in stake or delegator growth: \emph{when a validator changes an observable feature, is subsequent growth more or less likely?} We examine 2 types of events: commission changes and delinquency.

\emph{Commission changes.} We identify epochs in which a validator changes its commission rate. For each event, the outcome is evaluated only when the same validator has an observation at exactly epoch $t+K$. We distinguish between commission increases (fee hikes) and commission decreases (fee cuts). For each event, we evaluate whether the validator subsequently experiences stake growth or delegator growth after $K$ epochs. Events followed by another commission change or a delinquency-status change within the relevant horizon are excluded from the commission-change sample. We consider $K \in \{1,2,5,10,15,20\}$; because epochs represent approximately 2 days in our observation period, these values correspond to approximately 2, 4, 10, 20, 30 and 40 days, respectively. For each commission change event occurring at epoch $t$, we define the future outcome for stake growth as $Y_{i,t,K} = \mathbb{1}[S_{i,t+K} > S_{i,t}]$, and analogously for delegator growth as $Y_{i,t,K} = \mathbb{1}[D_{i,t+K} > D_{i,t}]$, where $D$ denotes the number of unique delegators. We then estimate the association between the type of commission change and the probability of subsequent growth. The resulting odds ratios indicate whether validators experiencing a fee hike or fee cut have higher or lower odds of growth under the reference category used in the model.

\emph{Delinquency and recovery.} We examine delinquency (non-performance) because it provides an observable signal of validator reliability. An outage event is defined as a transition from a non-delinquent state to a delinquent state. For each delinquency, we determine whether the validator has recovered after $K$ epochs or remains delinquent, distinguishing 2 outcomes among events with a valid observation at exactly epoch $t+K$: \emph{Recovered} (the validator is delinquent at the event epoch but is no longer delinquent at epoch $t+K$) and \emph{Persistent} (the validator remains delinquent at epoch $t+K$). We identify delinquency events as observed transitions from non-delinquent to delinquent status; validator status can change within an epoch. For each event at epoch $t$, we compare delegated stake and the number of unique delegators with their values at the exact epoch $t+K$, considering $K \in \{1,2,5,10,20\}$. We then examine whether recovery and persistent delinquency are associated with the probability of subsequent stake or delegator growth.

\para{Full Results}
\Cref{tab:study_a_static_stake,tab:study_a_static_delegators} report the static validator-level model for 2,060 validators, summarized in \Cref{sec:delegator_behavior}. \Cref{tab:study_b1} reports the commission-change results, and \Cref{tab:study_b2} the full delinquency-response results.

\begin{table}[tbhp]
\centering
\caption{Static validator features and probability of stake growth ($N=2{,}060$).}
\label{tab:study_a_static_stake}
\begin{tabular}{lrrrr}
\toprule
Feature & Coefficient & Std. Error & Odds Ratio & $p$-value \\
\midrule
Intercept & -0.8936 & 0.2009 & 0.4092 & $<0.001$ \\
Initial stake & -0.1715 & 0.0698 & 0.8424 & 0.014 \\
SFDP approved & +1.3357 & 0.1795 & 3.8027 & $<0.001$ \\
SFDP rejected & -1.6150 & 0.1693 & 0.1989 & $<0.001$ \\
SFDP retired & -0.6326 & 0.1832 & 0.5312 & $<0.001$ \\
Validator age & -0.3870 & 0.0648 & 0.6791 & $<0.001$ \\
Jito participation & +0.9094 & 0.1835 & 2.4829 & $<0.001$ \\
\bottomrule
\end{tabular}

\vspace{0.5em}
\begin{minipage}{0.95\textwidth}
\footnotesize
Notes: Infrastructure-location indicators (top 20 hosting cities, reference: other cities) are included as controls; full location-level coefficients are reported in \Cref{tab:appendix_location_stake}.
\end{minipage}
\end{table}

\begin{table}[thbp]
\centering
\caption{Static validator features and probability of delegator growth ($N=2{,}060$).}
\label{tab:study_a_static_delegators}
\begin{tabular}{lrrrr}
\toprule
Feature & Coefficient & Std. Error & Odds Ratio & $p$-value \\
\midrule
Intercept & -1.6725 & 0.2055 & 0.1878 & $<0.001$ \\
Initial stake & -0.0385 & 0.0501 & 0.9622 & 0.442 \\
SFDP approved/onboarded & +1.3114 & 0.1892 & 3.7115 & $<0.001$ \\
SFDP rejected & -0.2171 & 0.1578 & 0.8048 & 0.169 \\
SFDP retired & +0.3428 & 0.1842 & 1.4089 & 0.063 \\
Validator age & -0.5589 & 0.0603 & 0.5718 & $<0.001$ \\
Jito participation & +1.8839 & 0.1889 & 6.5791 & $<0.001$ \\
\bottomrule
\end{tabular}

\vspace{0.5em}
\begin{minipage}{0.95\textwidth}
\footnotesize
Notes: Infrastructure-location indicators (top 20 hosting cities, reference: other cities) are included as controls; full location-level coefficients are reported in \Cref{tab:appendix_location_delegators}.
\end{minipage}
\end{table}

\begin{table*}[htbp]
\centering
\caption{Commission changes and subsequent effect on validators across different $K$/epoch values. Baseline gives the probability of growth without a fee change. Coefficients are from logistic regressions, with odds ratios (OR) given by $\exp(\beta)$.}
\label{tab:study_b1}
\begin{tabular}{lrrrrrrrrrrr}
\toprule
& \multicolumn{5}{c}{Stake growth} & \multicolumn{5}{c}{Delegator growth} & \\
\cmidrule(lr){2-6} \cmidrule(lr){7-11}
$K$ & Baseline & \multicolumn{2}{c}{Fee hike} & \multicolumn{2}{c}{Fee cut} &
Baseline & \multicolumn{2}{c}{Fee hike} & \multicolumn{2}{c}{Fee cut} & $N$ \\
\cmidrule(lr){3-4} \cmidrule(lr){5-6} \cmidrule(lr){8-9} \cmidrule(lr){10-11}
& & $\beta$ & OR & $\beta$ & OR & & $\beta$ & OR & $\beta$ & OR & \\
\midrule
1  & 46.66 & -0.9005 & 0.4064 & 0.7667 & 2.1527 & 25.03 & -0.9997 & 0.3680 & -0.0976 & 0.9070 & 1,281 \\
2  & 48.52 & -0.5809 & 0.5594 & 0.5215 & 1.6845 & 29.77 & -0.8879 & 0.4115 & 0.0296 & 1.0301 & 1,011 \\
5  & 48.75 & -0.3224 & 0.7244 & 0.2723 & 1.3130 & 37.56 & -0.7356 & 0.4792 & 0.2275 & 1.2555 & 841 \\
10 & 51.16 & -0.4264 & 0.6529 & 0.4728 & 1.6045 & 43.22 & -0.7762 & 0.4602 & 0.5032 & 1.6540 & 738 \\
15 & 51.22 & -0.5849 & 0.5571 & 0.6336 & 1.8843 & 48.64 & -0.7478 & 0.4734 & 0.6934 & 2.0005 & 663 \\
20 & 57.22 & -0.7738 & 0.4612 & 1.0648 & 2.9004 & 53.22 & -0.9845 & 0.3736 & 1.1134 & 3.0447 & 594 \\
\bottomrule
\end{tabular}%
\end{table*}

\begin{table*}[htbp]
\centering
\small
\setlength{\tabcolsep}{4pt}%
\renewcommand{\arraystretch}{0.95}%
\caption{Validator delinquency and subsequent growth across different $K$/epoch values. Baseline gives the probability of growth without delinquency.}
\label{tab:study_b2}
\begin{tabular}{lrrrrrrrrrrr}
\toprule
& \multicolumn{5}{c}{Stake growth} & \multicolumn{5}{c}{Delegator growth} & \\
\cmidrule(lr){2-6} \cmidrule(lr){7-11}
$K$ & Baseline & \multicolumn{2}{c}{Recovered} & \multicolumn{2}{c}{Persistent} &
Baseline & \multicolumn{2}{c}{Recovered} & \multicolumn{2}{c}{Persistent} & $N$ \\
\cmidrule(lr){3-4} \cmidrule(lr){5-6} \cmidrule(lr){8-9} \cmidrule(lr){10-11}
& & $\beta$ & OR & $\beta$ & OR & & $\beta$ & OR & $\beta$ & OR & \\
\midrule
1  & 56.65 & 0.30 & 1.35 & -0.51 & 0.60 & 15.20 & -0.33 & 0.72 & -1.19 & 0.30 & 385/116 \\
2  & 52.60 & 0.04 & 1.04 & -0.89 & 0.41 & 19.99 & -0.11 & 0.90 & -1.81 & 0.16 & 377/51 \\
5  & 49.60 & 0.34 & 1.40 & -0.49 & 0.61 & 28.24 & 0.04 & 1.05 & -2.20 & 0.11 & 343/24 \\
10 & 49.26 & 0.74 & 2.10 & 0.03 & 1.03 & 35.97 & 0.16 & 1.17 & -1.37 & 0.25 & 313/16 \\
20 & 50.73 & 0.67 & 1.95 & -0.25 & 0.78 & 44.90 & 0.22 & 1.25 & -0.49 & 0.61 & 274/9 \\
\bottomrule
\end{tabular}%
\end{table*}

\para{Interpreting the Regression Models}
The purpose of the regression models in \Cref{sec:delegator_behavior} is to identify observable signals associated with validator attractiveness, rather than to claim that delegators explicitly inspect or respond to each feature.

For example, an odds ratio of 3 for a validator characteristic means that, conditional on the other variables in the model, validators possessing that characteristic have approximately 3 times the odds of experiencing the specified growth outcome relative to the reference condition. It does not mean that the characteristic causes 3 times as much stake to be delegated.

Similarly, a negative association between validator age and growth should not be interpreted as evidence that delegators dislike older validators. It indicates that, in the observed population and conditional on the variables included in the model, older validators are less likely to experience the defined growth outcome.

This distinction is particularly important for ecosystem-affiliation variables. For example, a strong association between SFDP participation and validator growth may reflect the visibility, reputation, delegation mechanisms or other characteristics associated with participating in the program. The analysis therefore identifies SFDP participation as a strong \emph{predictive signal} of validator attractiveness, but does not by itself establish which underlying mechanism produces the association.

\para{Full Location Coefficients}
\Cref{tab:appendix_location_stake,tab:appendix_location_delegators} report the full infrastructure-location coefficients for the static-feature models summarized in \Cref{sec:delegator_behavior} (\Cref{tab:study_a_static_stake,tab:study_a_static_delegators} there report only the non-location features).

\begin{table}[htbp]
\centering
\small
\setlength{\tabcolsep}{5pt}%
\renewcommand{\arraystretch}{0.95}%
\caption{Infrastructure-location coefficients, static model of stake growth ($N=2{,}060$).}
\label{tab:appendix_location_stake}
\begin{tabular}{lrrrr}
\toprule
City & Coefficient & Std. Error & Odds Ratio & $p$-value \\
\midrule
Frankfurt & +0.2603 & 0.1737 & 1.2973 & 0.134 \\
Amsterdam & -0.0115 & 0.1773 & 0.9886 & 0.948 \\
Chicago & +0.1110 & 0.2201 & 1.1174 & 0.614 \\
Moscow & -0.3922 & 0.3071 & 0.6756 & 0.202 \\
Ogden & +0.0947 & 0.3025 & 1.0993 & 0.754 \\
Vilnius & -0.2821 & 0.3117 & 0.7542 & 0.365 \\
London & +0.0926 & 0.2990 & 1.0970 & 0.757 \\
New York & +0.4104 & 0.2915 & 1.5075 & 0.159 \\
Oslo & +0.4810 & 0.3418 & 1.6178 & 0.159 \\
Berlin & +0.6513 & 0.3460 & 1.9180 & 0.060 \\
Singapore & +0.1346 & 0.3951 & 1.1440 & 0.733 \\
Tokyo & +0.6947 & 0.3807 & 2.0031 & 0.068 \\
Roubaix & +0.3112 & 0.4119 & 1.3651 & 0.450 \\
Los Angeles & -0.1155 & 0.4745 & 0.8909 & 0.808 \\
Ashburn & +0.3715 & 0.4195 & 1.4499 & 0.376 \\
Stockholm & +0.2667 & 0.4394 & 1.3056 & 0.544 \\
Strasbourg & -0.3370 & 0.5085 & 0.7139 & 0.508 \\
Warsaw & -0.8411 & 0.5859 & 0.4312 & 0.151 \\
Tirana & +0.0466 & 0.5149 & 1.0477 & 0.928 \\
Piscataway & -0.3655 & 0.5433 & 0.6939 & 0.501 \\
\bottomrule
\end{tabular}
\end{table}
\begin{table}[h]
\centering
\small
\setlength{\tabcolsep}{5pt}%
\renewcommand{\arraystretch}{0.95}%
\caption{Infrastructure-location coefficients, static model of delegator growth ($N=2{,}060$).}
\label{tab:appendix_location_delegators}
\begin{tabular}{lrrrr}
\toprule
City & Coefficient & Std. Error & Odds Ratio & $p$-value \\
\midrule
Frankfurt & +0.5874 & 0.1762 & 1.7993 & $<0.001$ \\
Amsterdam & +0.4394 & 0.1718 & 1.5517 & 0.011 \\
Chicago & -0.1071 & 0.2048 & 0.8984 & 0.601 \\
Moscow & -0.0208 & 0.2393 & 0.9794 & 0.931 \\
Ogden & +0.4037 & 0.2609 & 1.4974 & 0.122 \\
Vilnius & -0.1450 & 0.2645 & 0.8650 & 0.584 \\
London & +0.4077 & 0.2852 & 1.5034 & 0.153 \\
New York & +0.0405 & 0.2954 & 1.0413 & 0.891 \\
Oslo & +0.5528 & 0.3083 & 1.7380 & 0.073 \\
Berlin & +0.6219 & 0.3396 & 1.8624 & 0.067 \\
Singapore & +0.3946 & 0.4313 & 1.4837 & 0.360 \\
Tokyo & +0.8608 & 0.3992 & 2.3650 & 0.031 \\
Roubaix & +0.7065 & 0.3869 & 2.0269 & 0.068 \\
Los Angeles & +0.5438 & 0.3979 & 1.7225 & 0.172 \\
Ashburn & +0.6479 & 0.4104 & 1.9116 & 0.114 \\
Stockholm & -0.0823 & 0.4145 & 0.9210 & 0.843 \\
Strasbourg & -0.0105 & 0.4106 & 0.9895 & 0.980 \\
Warsaw & -0.2434 & 0.4413 & 0.7840 & 0.581 \\
Tirana & +0.5365 & 0.4340 & 1.7100 & 0.216 \\
Piscataway & +0.3303 & 0.4544 & 1.3914 & 0.467 \\
\bottomrule
\end{tabular}
\end{table}